\documentclass[journal,twoside,final]{IEEEtran}
\usepackage{mathrsfs}
\usepackage{amssymb}
\usepackage{amsfonts}
\usepackage{amsmath}
\usepackage{balance}
\usepackage{cuted}
\usepackage{psfrag}
\usepackage{graphicx}
\usepackage{tipa}
\usepackage{subfigure}
\usepackage{bigints}
\usepackage{cite}
\usepackage{xcolor}
\usepackage{array}
\usepackage{multirow}
\usepackage{soul}
\usepackage{epstopdf}

\begin{document}

\title{Optimizing Coverage in Convex Quadrilateral Regions with a Single UAV}

\author{Alexander~Vavoulas,~Nicholas~Vaiopoulos,~Konstantinos~K.~Delibasis,~and~Harilaos~G.~Sandalidis

\IEEEcompsocitemizethanks{
\IEEEcompsocthanksitem The authors are with the Department of Computer Science and Biomedical Informatics, University of Thessaly, Papasiopoulou 2-4, 35 131, Lamia, Greece. e-mails\{vavoulas,nvaio,kdelimpasis,sandalidis\}@dib.uth.gr.}
}

\maketitle

\begin{abstract}
 The integration of unmanned aerial vehicles (UAVs) into next-generation wireless networks has emerged as a promising solution for providing flexible and efficient coverage. This paper investigates the optimal deployment of a single UAV over an arbitrary convex quadrilateral region, employing a directional antenna with adjustable tilt that results in an elliptical ground coverage footprint. Two coverage scenarios are considered: (i) the largest inscribed ellipse, which maximizes coverage within the quadrilateral while excluding boundary regions, and (ii) the smallest circumscribed ellipse, which guarantees full coverage of the entire area.

An optimization framework is developed to determine the optimal UAV altitude by examining path loss, signal-to-noise ratio (SNR), and energy consumption. Based on a widely adopted path loss model, the altitude that minimizes the maximum path loss is derived, while the effect of antenna directivity on maximizing the minimum SNR at the coverage boundary is also analyzed. Furthermore, UAV energy consumption is evaluated by accounting for hovering, forward flight, and vertical take-off operations.

Numerical results illustrate the trade-offs among coverage efficiency, communication performance, and energy consumption under different propagation environments and antenna configurations. The analysis is further extended to all feasible inscribed and circumscribed ellipse configurations, providing a complete parametric characterization of the optimal altitude. In addition, a large-scale evaluation over randomly generated convex quadrilaterals offers a statistical assessment of the proposed framework and demonstrates its robustness under geometric variability. 
\end{abstract}


\section{Introduction}
\subsection{Preliminaries}
\IEEEPARstart{U}{nmanned} aerial vehicles (UAVs) are poised to play a pivotal role in the evolution of 6G networks, supporting ground base stations (BSs) and addressing the high-demand communication requirements of deployed sensor networks. They have witnessed remarkable significance in diverse sectors, including environmental monitoring, infrastructure inspection, disaster response, wildlife conservation, surveillance, and reconnaissance missions. The integration of UAVs is expected to significantly improve data rates and network capacity by capitalizing on their favorable communication links and adaptive mobility. However, extending the operational lifespan of UAVs and developing energy-efficient communication systems remain critical challenges for system designers and operators \cite{J:Geraci}, \cite{J:Vaiopoulos}. 

Given their unique operational characteristics, accurate air-to-ground propagation channel models are crucial for designing and evaluating UAV communication links, ensuring the reliable transmission of both control/non-payload data and payload data \cite{J:Khawaja}. In this context, trajectory planning plays a vital role in determining optimal flight paths, improving the efficiency of task execution, and allowing for the avoidance of obstacles \cite{J:Shukla}. Furthermore, optimizing the placement and altitude of UAVs, ideally using a single UAV, can significantly improve coverage and overall system performance. Effective coordination with BSs also contributes to the development of a flexible, integrated air-ground network \cite{J:Mozaffari}, \cite{J:Fotouhi}.

Previous studies have predominantly focused on optimizing deployment in a two-dimensional (2-D) space. However, in practical environments, UAVs operate in a three-dimensional (3-D) space, allowing them to maneuver vertically and navigate obstacles, including structures of varying heights, with greater efficiency. Effective placement procedures directly influence key performance metrics, including the probability of outage, deployment costs, quality of experience, and spectrum efficiency, ultimately improving overall network reliability. In addition, incorporating other critical system parameters is essential to ensure adaptability in dynamic or potentially hostile environments \cite{J:Carvajal-Rodriguez}.

\subsection{Related work}

Recent years have witnessed significant advances in the analysis and optimization of UAV-assisted wireless communication networks, with a strong emphasis on optimizing coverage, connectivity, and energy efficiency. Several comprehensive surveys and tutorials summarize the state-of-the-art in this field, offering broad overviews of UAV deployment strategies, system design considerations, and technical challenges~\cite{J:Carvajal-Rodriguez, J:Viet, J:Parvaresh, J:Ali, J:Elnabty}. Many of these works highlight that early research typically focused on the optimization of coverage and placement in regular geometric regions, such as circles and rectangles, mainly for mathematical tractability~\cite{Mozaffari2016,Zeng2016,Zeng2017,J:Alzenad}.

Within this context, 2-D models are commonly used to simplify analysis, plan UAV routes, and coverage areas as flat planes that neglect 3-D obstacles. For example, Wu~\emph{et al.}~\cite{J:Wu} developed a 2-D trajectory model for UAV communication networks, and Mardani~\emph{et al.}~\cite{J:Mardani} leveraged similar models to improve communication quality and support seamless video transmission. More recent studies have begun to incorporate practical considerations beyond idealized 2-D domains, including 3-D trajectory optimization~\cite{Zhang2019}, cooperative coverage with ground BSs~\cite{J:Lyu}, dynamic user association strategies~\cite{J:Cheng}, and the impact of channel characteristics such as fading and shadowing~\cite{J:Cattai,J:Yang21}. In addition, issues related to deployment in heterogeneous environments and energy-efficient operation have attracted growing attention~\cite{Xie2021,Mozaffari2017}.  Moreover, several line-of-sight (LoS) probability models have been proposed in the literature; however, many rely on simplified Manhattan-grid assumptions and ITU-defined built-up parameters, which may not fully capture the randomness of real urban environments. More advanced approaches address these limitations by incorporating more realistic city layouts with varying building sizes and shapes \cite{J:Saboor23, C:Saboor25, J:Mohammed, J:Gapeyenko, J:Gholami, J:Yang, J:Saoud}.

A substantial body of work has specifically addressed UAV altitude optimization. For example, Al-Hourani~\cite{J:Al-Hourani} examined the optimal altitude of UAVs to maximize the coverage of circular ground regions, while Lyu~\emph{et al.}~\cite{J:Lyu} analyzed fixed-altitude scenarios to minimize the number of required coverage disks. Alzenad~\emph{et al.}~\cite{J:Alzenad} proposed efficient UAV placement strategies to enhance user coverage while minimizing transmit power. Further studies have explored the effects of Rician fading on optimal altitude and coverage~\cite{C:Azari, J:Liu}, the joint optimization of altitude and 2-D positioning~\cite{J:Alzenad2}, and the role of directional antennas in multi-tier UAV networks~\cite{J:Zhang}. Zhou~\emph{et al.}~\cite{J:Zhou} also addressed UAV positioning and orientation to minimize deployment costs in irregular ground coverage scenarios.

Despite this progress, coverage analysis and optimization for arbitrary and irregular regions remain relatively underexplored. Although some recent studies have tackled challenges in time-varying coverage shapes~\cite{J:Xie2024,C:Srivastava}, a comprehensive analytical framework for the deployment of UAVs over arbitrary convex quadrilaterals is still lacking. The present work addresses this gap by systematically treating coverage and altitude optimization in such regions and further distinguishes itself by incorporating geometric decomposition and advanced channel modeling to enhance practical applicability. 

\subsection{Motivation}

Previous studies predominantly assumed circular coverage areas, where the antenna radiation pattern is perpendicular to the ground. However, tilting the antenna relative to this perpendicular axis results in an elliptical coverage footprint. Such a footprint provides a substantially better geometric match than a circular one, reducing coverage gaps or excessive overlap outside the target area. This improved shape adaptability enables more efficient utilization of UAV resources and more accurate modeling of realistic deployments.
In this context, \cite{J:You} proposed an energy-efficient 3-D UAV placement algorithm that incorporates antenna tilt to generate elliptical coverage on the ground. Moreover, in our previous work, we considered elliptical coverage footprints in scenarios where terminal positions are randomly distributed, further emphasizing the importance of using ellipses \cite{J:Vavoulas}.

In real-world applications, the target coverage area is often an arbitrary convex quadrilateral rather than an idealized circular or square region. Thus, quadrilateral coverage modeling provides a more realistic representation for elongated or irregularly shaped environments with non-uniform user distributions, such as cultural or sporting events, urban landscapes with obstacles, and precision agriculture \cite{J:Mukhamediev}. The inherent geometric complexity, which arises from the variable side lengths, angles, and absence of symmetry, complicates both the modeling of coverage footprints and the optimization of UAV placement. These factors directly affect the characterization of key performance metrics, such as path loss, SNR, and energy consumption, which impact  battery life and the duration of UAV operations. 
Consequently, determining the optimal altitude must balance communication performance and power efficiency to maximize UAV operational effectiveness.
As a result, optimizing UAV altitude over convex quadrilateral regions requires more advanced mathematical techniques and careful geometric analysis and generally precludes the use of simplified closed-form solutions available for symmetric domains. 

Although real-world coverage regions are occasionally represented by convex $n$-sided polygons, such shapes introduce significant analytical complexity. In fact, it is possible to determine the inscribed and circumscribed ellipses for convex polygons~\cite{B:Boyd}, but these solutions are typically tractable only for certain classes of polygons. In contrast, convex quadrilaterals provide a practical balance between modeling flexibility and analytical simplicity, enabling efficient coverage optimization. 
Moreover, any arbitrary convex polygon can be systematically partitioned into a set of quadrilateral subregions~\cite{C:MullerHannemann, C:Lubiw}, which facilitates the scalability of the proposed framework to more complex domains.  

It is noted that \cite{J:Vavoulas2} investigates the deployment of multiple UAVs in large convex square areas as an extension of the current single UAV framework. The number of UAVs required for full coverage depends on system-level constraints such as data rate, error probability, and energy efficiency, which may differ even for identical geographical areas. Therefore, \cite{J:Vavoulas2} is not an alternative but a complementary study, together with the present work, which forms part of a broader framework for analyzing UAV coverage in wireless networks.
  
\subsection{Contribution}
This study focuses on covering a convex quadrilateral region with a single UAV, where the elliptical coverage footprint of the UAV is adjusted by altitude and antenna tilt/semi-apex angles to match the shape and size of the designated area as closely as possible. 
This deployment setting is relevant in practical applications such as localized events or critical infrastructure monitoring, where the area of interest is sufficiently compact to be served by one UAV.  

The main contributions of this work are:
\begin{itemize}
    \item The ellipse is interpreted as the ground footprint of an oblique circular cone corresponding to an antenna-generated iso-surface, thereby establishing a physically grounded geometric formulation. Within this framework, geometric constructions for the maximum-area inscribed ellipse and the minimum-area circumscribed ellipse of an arbitrary quadrilateral are employed.
    
    \item The proposed framework enables rapid evaluation of quadrilateral coverage scenarios without the need for case-specific simulations while also providing analytical insight into the interplay among ellipse geometry, propagation conditions, and coverage performance metrics, thereby facilitating system design and optimization.
    
    \item The methodology is sufficiently general to accommodate multiple propagation models associated with different urban densities, building dimensions and spacing, and obstacle configurations, thus providing a flexible baseline framework for practical network planning applications.
    
    \item The proposed framework can be naturally extended to arbitrary polygonal regions through quadrilateral decomposition, enabling scalable analysis of complex coverage areas under diverse propagation environments and planning constraints. Moreover, the proposed method exhibits marginally improved performance compared with the previously reported homography-based approach in~\cite{J:Vavoulas2}.
\end{itemize}
In addition to coverage optimization, the present study investigates the determination of the optimal UAV operating altitude by jointly considering the following key performance factors:
\begin{itemize}
    \item \textbf{Path loss:} A widely adopted air-to-ground propagation model incorporating both LoS and non-line-of-sight (NLoS) components is employed to numerically evaluate the maximum path loss and its effect on coverage performance.
    \item \textbf{SNR:} Based on the considered propagation model, a directional antenna radiation pattern is incorporated to estimate the minimum SNR at the coverage boundary, thereby providing a more accurate characterization of communication quality.
    \item \textbf{Total energy consumption:} UAV energy efficiency is analyzed by accounting for multiple flight regimes, including hovering, forward flight, and vertical takeoff, in order to determine the operating altitude that minimizes total energy expenditure.
\end{itemize}
We further conduct a large-scale statistical analysis over randomly generated convex quadrilaterals to demonstrate the robustness and generality of the proposed framework. In contrast to purely simulation-driven approaches, the presented methodology provides reusable analytical tools for rapid system-level evaluation across a wide range of deployment scenarios and establishes a flexible foundation for subsequent optimization and learning-based extensions. 

\subsection{Organization}
The remainder of the paper is structured as follows. Section II introduces the system model, describing the UAV placement strategy and the convex quadrilateral coverage area. It also formulates the steps to determine the optimal inscribed and circumscribed ellipses. Section III focuses on UAV altitude optimization. It first analyzes the path loss to determine the altitude that minimizes the maximum path loss. Then, it examines the impact of antenna directivity on SNR to identify the altitude that maximizes the minimum SNR. Finally, it explores energy consumption by obtaining the altitude that minimizes total power consumption. Section IV presents a case study with numerical results in different environments and antenna configurations, illustrating the trade-offs between coverage, SNR, and energy consumption. Section V concludes the paper by summarizing key findings and discussing future research directions, including dynamic mobility and multi-UAV coordination.

\section{Elliptical Cell Generation}

\subsection{Setup configuration}

\begin{figure}[h]  
    \centering  
    \subfigure[Inscribed ellipse]{\includegraphics[width=0.8\columnwidth]{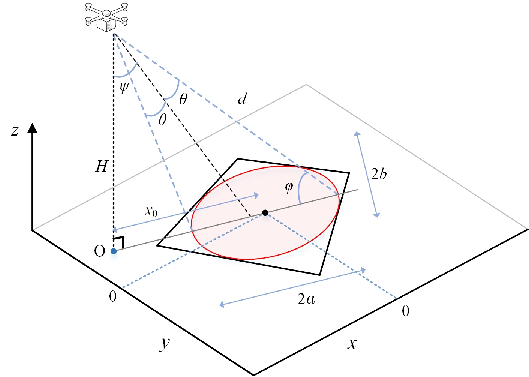}}  
    \hspace{0in}  
    \subfigure[Circumscribed ellipse]{\includegraphics[width=0.8\columnwidth]{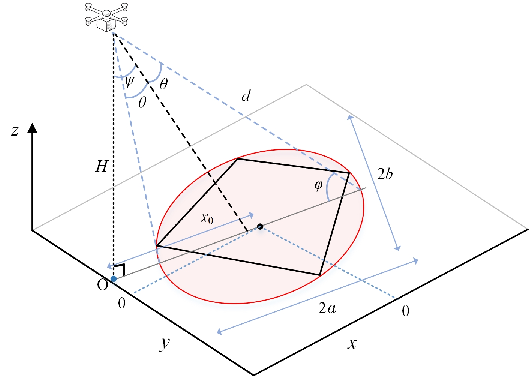}}  
    \hspace{0in}  
    \caption{System configurations}  
    \label{Figure1}  
\end{figure}

In this study, we examine two distinct scenarios:

\begin{itemize}
    \item \textbf{Scenario 1:} In this case, the goal is to cover the interior of the quadrilateral, excluding its boundary, as depicted in Fig. \ref{Figure1}(a). The largest inscribed ellipse has been identified as a good fit for this objective.
    \item \textbf{Scenario 2:} Here, the objective shifts to full coverage of the quadrilateral, including its boundaries, as indicated in Fig. \ref{Figure1}(b). This requires the smallest circumscribed ellipse that encompasses the entire region.
\end{itemize}

These two coverage scenarios correspond to different practical deployment requirements in UAV-assisted communication systems. The inscribed ellipse scenario is suitable for applications where reliable service is required primarily within the interior of the region, while boundary areas may be unoccupied, inaccessible, or intentionally excluded to avoid interference with neighboring cells. Examples include precision agriculture monitoring, environmental sensing over hazardous terrain, and coverage of localized crowds positioned away from physical barriers.

The circumscribed ellipse scenario addresses applications that demand complete coverage of the entire region, including its boundaries. Such requirements commonly arise in emergency response operations, disaster recovery, public safety surveillance, border monitoring, and temporary network provisioning for large outdoor events, where communication availability must be guaranteed across the whole area.

For clarity, we first summarize the main geometric quantities used in the coverage model.  The ground region is defined on the Cartesian plane $\{x,y\}$, while the UAV operates in 3-D space at an altitude $H$ \cite{C:Tang}. The UAV carries a directional antenna whose  emission axis is tilted by an angle $\psi$ with respect to the ground normal and forms a cone with a semi-apex angle $\theta$. The intersection of this cone with the ground plane  produces an elliptical footprint characterized by semi-axes $a$ and $b$, center $C=(C_x,C_y)$, orientation angle $\Theta$, and a planar offset between the UAV ground  projection and the ellipse center. These parameters determine the size, position, and orientation of the coverage region and will be used in the analytical formulation that follows.

In the following, we consider the optimization of the coverage of a ground area enclosed by an arbitrary convex quadrilateral $Q$ in the $\{x, y\}$ plane, with vertices denoted $P_i = (x_i, y_i)$ for $i = 1, \dots, 4$. 
The goal is to provide optimal coverage of the area $Q$ using a single UAV operating at different altitudes $H$. The UAV is equipped with a directional antenna that exhibits angle-dependent gain, as discussed in Section III. 
The antenna angular characteristics are defined by two parameters, the semi-apex angle $\theta$ and the tilt angle $\psi$ of the emission axis with respect to the normal ground vector. The antenna is assumed to have identical half-power beamwidths in both azimuth and elevation, resulting in a circular ground footprint when the tilt angle is zero. 

Let us denote the major and minor axes of the elliptical footprint (ground radiation pattern) by $2a$ and $2b$, respectively. Figure \ref{Figure1} depicts this geometry setting.
The UAV's ground projection, represented by the point $O$, lies on the major axis of the ellipse generated by the antenna's radiation pattern, as seen from the UAV's position at altitude $H$.  
This occurs because, when the cone axis is tilted, its intersection with the ground plane becomes an ellipse, with its major axis aligned along the plane defined by the cone axis and the normal to the ground plane. Consequently, due to the cone’s rotational symmetry, the UAV’s ground projection lies along the major axis of this ellipse.
Allowing the projection to deviate from the major axis would require elliptic cone emission, thereby breaking the geometric coupling among altitude, tilt angle, and the ellipse parameters.
 
The angle between the emission axis and any point of the ellipse is constant and equal to the semi-apex angle $\theta$. It can be easily verified that moving the UAV to the other side of the center of the ellipse with the same $x_0,\theta \text{ and } \psi$ will generate a symmetric setup with an identical elliptical footprint.

The planar offset, $x_0$, between the projected position of the UAV $O$ and the center of the ellipse is determined by the relative angles $\theta$ and $\psi$. 
\begin{equation}
  x_0=a - \text{sign}(\theta-\psi) H\tan(\theta-\psi),  
\end{equation}
where the sign of $(\theta-\psi)$ determines whether $O$ lies inside or outside the ellipse (positive and negative, respectively). Specifically, when $O$ lies within the ellipse, we have $\psi \leq \theta$; whereas, if $O$ lies outside the ellipse, we have $\psi > \theta$. The coordinates of the position of the UAV $O_x, O_y, H$ are defined in Section II.D as a function of the ellipse center $C_x, C_y$, the inclination of the major axis, and the planar offset $x_0$.

Moreover, for a given value of $H$, the angles $\psi$ and $\theta$ are expressed in terms of $a$ and $b$  as

\begin{equation}
    \psi=\arccos\left(\frac{\sqrt{b^2 H^2+b^4}}{\sqrt{a^2 H^2+b^4}}\right),
    \label{psi}
\end{equation}
\begin{equation}
    \theta=\arcsin\left(\frac{b^2}{\sqrt{a^2 H^2+b^4}}\right).
    \label{theta}
\end{equation}
This formulation enables a precise calculation of the UAV coverage area at different altitudes, providing a foundation for optimal coverage strategies for the convex quadrilateral $Q$.  It should be noted that the planar offset $x_0$ of the UAV is not determined independently of its altitude $H$. 
As demonstrated previously, $x_0$ is a function of $H$, the antenna tilt angle $\psi$, and the semi-apex angle $\theta$, both of which are themselves dependent on $H$. Consequently, the planar offset and altitude are inherently coupled through geometric and communication constraints. 
This unified treatment ensures that the proposed framework optimizes the location and altitude of the UAV, rather than addressing these parameters separately or in a fragmented way.

\subsection{Inscribed ellipse with largest area}
\begin{figure}[h]
\centering
\subfigure[]{\includegraphics[width=0.8\columnwidth]{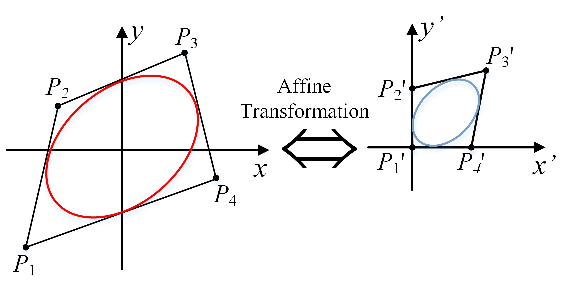}}
\hspace{0.1in} 
\subfigure[]{\includegraphics[width=0.85\columnwidth]{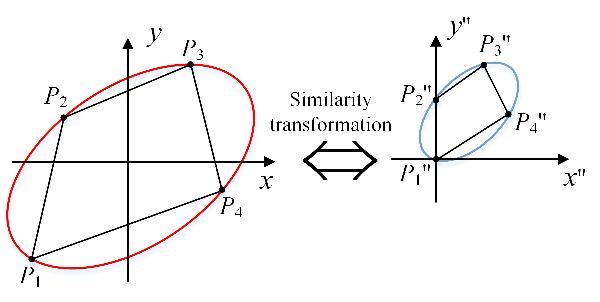}} 
\hspace{0in} 
\caption{Geometric transformations for determining the inscribed and circumscribed ellipse.}
\label{Figure2}
\end{figure}

To determine the ellipse with the maximal area inscribed in $Q$, we follow the step-by-step methodology in \cite[sec. I]{B:Horwitz}. More precisely, let $\mathbf{T_a}$ be the affine transformation that maps $Q$ to the quadrilateral
$Q'$ with vertices $P'_1=(0,0)$, $P'_2=(0,1)$, $P'_3=(s,t)$, and $P'_4=(1,0)$ as depicted in Fig. \ref{Figure2}(a).
$\mathbf{T_{a}}$ applies to any point $(x_i,y_i)$ according to
\begin{gather}
 \begin{bmatrix} x_i'  \\
y_i'\\
1 \end{bmatrix}
 =\mathbf{T_a}
 \begin{bmatrix} x  \\
y\\
1 \end{bmatrix}
=
  \begin{bmatrix}
   \rho_{11} & \rho_{12} & \rho_{13} \\
   \rho_{21} & \rho_{22} & \rho_{23} \\
   0 & 0 & 1
   \end{bmatrix}
 \begin{bmatrix} x_i \\
y_i\\
1 \end{bmatrix},
\label{AffTrans}
\end{gather}
with $i=1,...,4$. Using the coordinates of the vertices of $Q$ and the points $P'_1$, $P'_2$, and $P'_4$, $\mathbf{T_a}$ yields, by solving two systems of linear equations in \eqref{AffTrans} for $i=1,2,4$\footnote{The first system of three equations computes the unknowns $\rho_{11}, \rho_{12}, \rho_{13}$, and the second one determines $\rho_{21}, \rho_{22}, \rho_{23}$.}. Then, the point $P'_3$ is computed by applying $\mathbf{T_a}$ to $P_3$, so $Q'$ is fully defined.
The general equation representing all inscribed ellipses in the coordinate system $(x', y')$ is given by \cite[eq. (1.5)]{B:Horwitz}
\begin{equation}
    \mathcal{I}_1x'^2\hspace{-2pt}+\hspace{-2pt}\mathcal{I}_2(q)x'y'\hspace{-2pt}+\hspace{-2pt}\mathcal{I}_3(q)y'^2\hspace{-2pt}+\hspace{-2pt}\mathcal{I}_4(q)x'\hspace{-2pt}+\hspace{-2pt}\mathcal{I}_5(q)y'\hspace{-2pt}+\hspace{-2pt}\mathcal{I}_6(q)\hspace{-2pt}=\hspace{-2pt}0,
\label{InscEl}
\end{equation}
where $\mathcal{I}_1=t^2$, $\mathcal{I}_2(q)=4q^2(s-1)t+2qt(s-t+2)-2st$, $\mathcal{I}_3(q)=((1-q)s+qt)^2$, $\mathcal{I}_4(q)=-2qt^2$, $\mathcal{I}_5(q)=-2qt((1-q)s+qt)$, $\mathcal{I}_6(q)=q^2t^2$, with parameter $q\in[0,1]$. Among the possible values of $q$, the one that maximizes the inscribed elliptical region is determined as \cite[ch. 7.2]{B:Horwitz}
\begin{equation}
q=\frac{-(st\hspace{-3pt}-\hspace{-3pt}t\hspace{-3pt}+\hspace{-3pt}1)\hspace{-3pt}+\hspace{-3pt}\sqrt{(st\hspace{-3pt}-\hspace{-3pt}t\hspace{-3pt}+\hspace{-3pt}1)^2+t(s\hspace{-3pt}-\hspace{-3pt}1)(t\hspace{-3pt}-\hspace{-3pt}s\hspace{-3pt}+\hspace{-3pt}2)}}{(t-1)(t-s+2)},
\label{qval}
\end{equation}
The unique ellipse of the maximal area inscribed in $Q$ can be expressed in quadratic form as
\begin{equation}
\mathcal{B}_1x^2+\mathcal{B}_2xy+\mathcal{B}_3y^2+\mathcal{B}_4x+\mathcal{B}_5y+\mathcal{B}_6=0,
\label{InscElFinal}
\end{equation}
where $\mathcal{B}_1,...,\mathcal{B}_6\in \mathbb{R}$ can be easily determined by substituting \eqref{qval} into \eqref{InscEl} and applying the inverse affine transform, as stated in \eqref{AffTrans}.

\subsection{Circumscribed ellipse with smallest area}
In this scenario, a similarity transformation $\mathbf{T}_s$ maps $Q$ to the quadrilateral $Q''$ in the $\{x'', y''\}$ plane, with vertices $P''_1=(0,0)$, $P''_2=(0,1)$, $P''_3=(s,t)$, and $P''_4=(v,w)$ as depicted in Fig. \ref{Figure2}(b) according to
\begin{gather}
 \begin{bmatrix} x_i''  \\
y_i''\\
1 \end{bmatrix}
 =\mathbf{T}_s
 \begin{bmatrix} x_i  \\
y_i\\
1 \end{bmatrix}
=
  \begin{bmatrix}
   \sigma \cos(\phi) & -\sigma\sin(\phi) & \delta x \\
   \sigma\sin(\phi) & \sigma \cos(\phi) & \delta y \\
   0 & 0 & 1
   \end{bmatrix}
 \begin{bmatrix} x_i  \\
y_i\\
1 \end{bmatrix},
\label{IsoTrans}
\end{gather}
where $(x_1'',y_1'')=(0,0)$, $(x_2'',y_2'')=(0,\mathrm{d})$, and $\mathrm{d}$ is the Euclidean distance between $P_1$ and $P_2$. Using the coordinates of the vertices of $Q$ and the points $P''_1$ and $P''_2$, we can determine the required (signed) angle: $\phi=\frac{\pi}{2}- \arctan\left(\frac{y_2-y_1}{x_2-x_1} \right )$ if $x_2-x_1 \ne 0$, else $\phi=0.$ The required translations in (\ref{IsoTrans}) are easily derived as $\delta x=-x_1$ and $\delta_y=-y_1$. Finally, the scale factor $\sigma$ is set to $\sigma=\frac{1}{\mathrm{d}}$. Once $\mathbf{T}_s$ is calculated, we can find $P''_3$ and $P''_4$, which define $Q''$ \cite[sec. II]{B:Horwitz}. 

The general equation representing all circumscribed ellipses in the coordinate system $(x",y")$ is given by \cite[eq. (8.1)]{B:Horwitz}
\begin{equation}
    \mathcal{C}_1(u)x''^2\hspace{-2pt}+\hspace{-2pt}\mathcal{C}_2(u)x''y''\hspace{-2pt}+\hspace{-2pt}\mathcal{C}_3(u)y''^2\hspace{-2pt}+\hspace{-2pt}\mathcal{C}_4(u)x''\hspace{-2pt}+\hspace{-2pt}\mathcal{C}_5(u)y''\hspace{-2pt}\hspace{-2pt}=\hspace{-2pt}0,
\label{CircEl}
\end{equation}
where $\mathcal{C}_1(u)=sv(t-w)u$, $\mathcal{C}_2(u)=(sw^2-t^2v+vt-ws-sv(s-v)u)$, $\mathcal{C}_3(u)=sv(t-w)$, $\mathcal{C}_4(u)=(v(t-1)+(1-w)s)tw-(vt-ws)svu$, and $\mathcal{C}_5(u)=-sv(t-w)$. Among the possible values of $u$, the value that minimizes the circumscribed elliptical region is given by \cite[ch. 11.1]{B:Horwitz}
\begin{equation}
\begin{split}
s^3(s-1)^2u^3+s^2t(2(s-1)^2+st+s+t-1)u^2- \\
st^2(2(t-1)^2+st+s+t-1)u-t^3(t-1)^2=0,
\label{uval}
\end{split}
\end{equation}

The unique ellipse of the minimal area circumscribed about $Q$ is obtained as
\begin{equation}
\mathcal{D}_1x^2+\mathcal{D}_2xy+\mathcal{D}_3y^2+\mathcal{D}_4x+\mathcal{D}_5y+\mathcal{D}_6=0,
\label{CircElFinal}
\end{equation}
where $\mathcal{D}_1,...,\mathcal{D}_6\in \mathbb{R}$ can be easily determined by substituting the root of \eqref{uval} into \eqref{CircEl} and applying the inverse similarity transform $\mathbf{T}_s^{-1}$.

\subsection{Geometric relationships}
The major and minor semi-axes, $a$ and $b$, of the inscribed and circumscribed ellipses can be expressed through \eqref{InscElFinal} and \eqref{CircElFinal}, respectively, as
\begin{equation}
    a=\sqrt{\mu\frac{\mathcal{F}_1+\mathcal{F}_3+\sqrt{(\mathcal{F}_1-\mathcal{F}_3)^2+\mathcal{F}_2^2}}{2}},
\label{alpha}
\end{equation}
\begin{equation}
    b=\sqrt{\mu\frac{\mathcal{F}_1+\mathcal{F}_3-\sqrt{(\mathcal{F}_1-\mathcal{F}_3)^2+\mathcal{F}_2^2}}{2}},
\label{beta}
\end{equation}
where $\mu=4\delta_1\delta_2^{-2}$, $\delta_1=\mathcal{F}_3\mathcal{F}_4^2+\mathcal{F}_1\mathcal{F}_5^2-\mathcal{F}_2\mathcal{F}_4\mathcal{F}_5-\mathcal{F}_6\delta_2$, $\delta_2=4\mathcal{F}_1\mathcal{F}_3-\mathcal{F}_2^2$, while $\mathcal{F}_i = \mathcal{B}_i$ (inscribed) and $\mathcal{F}_i = \mathcal{D}_i$ (circumscribed) for $i = 1, \ldots, 6$.

The coordinates of the center of the ellipse and the inclination of the major axis $\varTheta$ are given as follows:
\begin{equation}
\mathsf{C}_x=\frac{\mathcal{F}_2\mathcal{F}_5-2\mathcal{F}_3\mathcal{F}_4}{4\mathcal{F}_1\mathcal{F}_3-\mathcal{F}_2^2},
\label{Cx}
\end{equation}
\begin{equation}
\mathsf{C}_y=\frac{\mathcal{F}_2\mathcal{F}_4-2\mathcal{F}_1\mathcal{F}_5}{4\mathcal{F}_1\mathcal{F}_3-\mathcal{F}_2^2},
\label{Cy}
\end{equation}
\begin{equation}
\varTheta=\frac{1}{2}\text{arccot}\left(\frac{\mathcal{F}_1-\mathcal{F}_3}{\mathcal{F}_2}\right)+\frac{\pi}{2},
\end{equation}
The projected UAV position (on the ground) is given by
\begin{align}
    O_x&=\mathsf{C}_x-x_0\cos(\varTheta)\\
    O_y&=\mathsf{C}_y-x_0\sin(\varTheta)
\end{align}
The position of the UAV in 3-D space is defined by $(O_x,O_y,H).$

\section{UAV Placement and Altitude Optimization}
This section explores the ideal UAV altitude required to cover both inscribed and circumscribed ellipses determined in the previous section, taking into account path loss, SNR, and energy consumption.

\subsection{Optimal altitude vs. path loss}
A simplified path loss model is used to calculate the maximum path loss, where the receiver (Rx) has a certain probability of maintaining a LoS connection with the UAV. This probability is affected by environmental factors and the altitude of the UAV, as detailed in \cite{J:Al-Hourani}.  The adopted propagation model is among the most widely used in the literature, and its validity has been extensively demonstrated; although different propagation models may yield different numerical values for the optimal altitude due to their underlying assumptions and parameterizations, the qualitative behavior remains consistent, particularly in terms of the dependence of the optimal altitude on coverage requirements, footprint geometry, and environmental conditions.  The LoS probability is given by
\begin{equation}
\mathbb{P}(\text{LoS})=\frac{1}{1+\eta\exp(-\kappa(\varphi-\eta))},
    \label{PLOS}
\end{equation}
where $\eta$ and $\kappa$ are parameters of the sigmoid function that depend on the environment, and $\varphi$ represents the angle in degrees corresponding to the boundary of the coverage area given by the expression
\begin{equation}
\varphi=\left\{ 
\begin{array}{l}
\arctan\left(\frac{H}{2a-H\tan(\theta-\psi)}\right),\text{ \ }\psi\leq \theta \\ 
\\ 
\arctan\left(\frac{H}{2a+H\tan(\psi-\theta)}\right),\text{ }\psi>\theta, \\ 
\end{array}%
\right.   \label{phi}
\end{equation}
The NLoS probability is given by 
\begin{equation}
 \mathbb{P}(\text{NLoS})=1-\mathbb{P}(\text{LoS}). 
\end{equation}

The maximum path loss (in dB) between the UAV and a ground Rx occurs at the boundary of the coverage area, i.e., \cite{J:Al-Hourani}
\begin{equation}
PL_{\text{max}} = \mathbb{P}(\text{LoS}) \cdot PL_{\text{LoS}} + \mathbb{P}(\text{NLoS}) \cdot PL_{\text{NLoS}},
    \label{PathLoss}
\end{equation}
where 
\begin{align}
\begin{split}
    PL_{\text{LoS}}= 20\log d+20\log\left(\frac{4\pi f}{c}\right)+\xi_{\text{LoS}}\\
    PL_{\text{NLoS}}=20\log d+20\log\left(\frac{4\pi f}{c}\right)+\xi_{\text{NLoS}},
     \label{PathLoss2}
\end{split}
\end{align}
 $PL_{\text{LoS}}$ and $PL_{\text{NLoS}}$ denote the path losses under LoS/NLoS conditions, respectively, $d=\sqrt{H^{2}+(x_{0}+a)^2}$ represents the distance from the UAV to the boundary of the coverage area, $f$ is the operating frequency, and $\xi_{\text{LoS}}$, $\xi_{\text{NLoS}}$ are the mean values of excessive losses due to scattering and shadowing.

Multiple sets of values ${H, \theta, \psi}$ can generate inscribed or circumscribed elliptical footprints, as defined by \eqref{InscElFinal} and \eqref{CircElFinal}, or equivalently by \eqref{alpha} and \eqref{beta}. 
However, among these configurations, only one configuration minimizes $PL_{\text{max}}$. The goal is to identify the optimal altitude, $H_{\text{OPT}}$, at which $PL_{\text{max}}$ is minimized. To achieve this, we first express the angles $\psi$ and $\theta$ in \eqref{phi} using \eqref{psi} and \eqref{theta}, and substitute them into \eqref{PLOS}. 
Finally, through straightforward algebraic manipulations of \eqref{PLOS}, \eqref{PathLoss}, and \eqref{PathLoss2}, the following unified expression is derived

\begin{equation}
\small 
\begin{aligned}
PL_{\text{max}}(H) = & \, \frac{\xi_{\text{LoS}} - \xi_{\text{NLoS}}}{1\hspace{-2pt} +\hspace{-2pt} \eta \exp \left(\hspace{-2pt} -\kappa \left( \arctan \left( \frac{Hb}{ab + \sqrt{(b^2 + H^2)(a^2 - b^2)}} \right)\hspace{-2pt} -\hspace{-2pt} \eta \right) \right)} \\
& + 10 \log \left( H^2 + \left( \frac{ab + \sqrt{(b^2 + H^2)(a^2 - b^2)}}{b} \right)^2 \right) \\
& + 20 \log \left( 4 \pi f / c \right) + \xi_{\text{NLoS}}.
\label{PLmax}
\end{aligned}
\end{equation}

The optimal altitude, $H_{\text{OPT}}$, is determined by identifying the value $H$ that sets the first derivative of $PL_{\text{max}}$ with respect to $H$ equal to zero, that is,
\begin{equation}
\frac{\partial PL_{\text{max}}(H)}{\partial H}=0.
    \label{Der1}
\end{equation}
This process results in the nonlinear equation of \eqref{Eq1} which has no closed-form solution and is solved using numerical methods\footnote{The optimal altitude is determined by the environmental conditions, which in turn enables the calculation of the corresponding optimal angles, $\psi$ and $\theta$, using \eqref{psi} and \eqref{theta}.} \\ \hrule

\begin{strip}
\footnotesize

\begin{equation}
\frac{{9{C_1}\eta \kappa \left( {a\sqrt {{b^2} + H_{{\rm{OPT}}}^2}  + {b^2}{C_2}} \right)\exp (\kappa \eta ){C_3}({H_{{\rm{OPT}}}})}}{{\pi \sqrt {{b^2} + H_{{\rm{OPT}}}^2} \left( {{{\left( {a + {C_2}\sqrt {{b^2} + H_{{\rm{OPT}}}^2} } \right)}^2} + H_{{\rm{OPT}}}^2} \right){{\left( {{C_3}({H_{{\rm{OPT}}}}) + \eta \exp (\eta \kappa )} \right)}^2}}} + \frac{{{H_{{\rm{OPT}}}}\left( {\frac{{a{C_2}}}{{\sqrt {{b^2} + H_{{\rm{OPT}}}^2} }} + C_2^2 + 1} \right)}}{{\ln (10)\left( {{{\left( {a + {C_2}\sqrt {{b^2} + H_{{\rm{OPT}}}^2} } \right)}^2} + H_{{\rm{OPT}}}^2} \right)}} = 0,
\label{Eq1}
\end{equation}
\end{strip}

\hrule
\noindent In \eqref{Eq1} we denote $C_{1}=\xi_{\text{LoS}} - \xi_{\text{NLoS}}$, $C_{2}=\sqrt{a^2-b^2}/b$ and $C_{3}(H)=\exp\left(\frac{180}{\pi}\arctan\left(\frac{H}{a+C_{2}\sqrt{b^2+H^2}}\right)\right)$. It must be noted that the path loss function $PL_{\rm max}(H)$ is continuous for all feasible $H>0$ as it is composed of elementary continuous functions with strictly positive denominators and arguments over the feasible range. Furthermore, Appendix A summarizes the proof that $PL_{\max}(H)$ is strictly unimodal.  

\subsection{Optimal altitude vs. SNR}
Analyzing the relationship between altitude and maximum path loss provides an initial approximation to optimize UAV positioning. However, a more thorough and precise analysis requires the inclusion of the antenna radiation pattern, as it plays a vital role in the link budget calculation and offers a deeper understanding of the optimization process. In this regard, the optimal altitude, $H_{\text{OPT*}}$, at which the minimum SNR, $\gamma_{\text{min}}$, is maximized, will be determined.

\begin{figure}[h]
\centering
\includegraphics[keepaspectratio,width=3in]{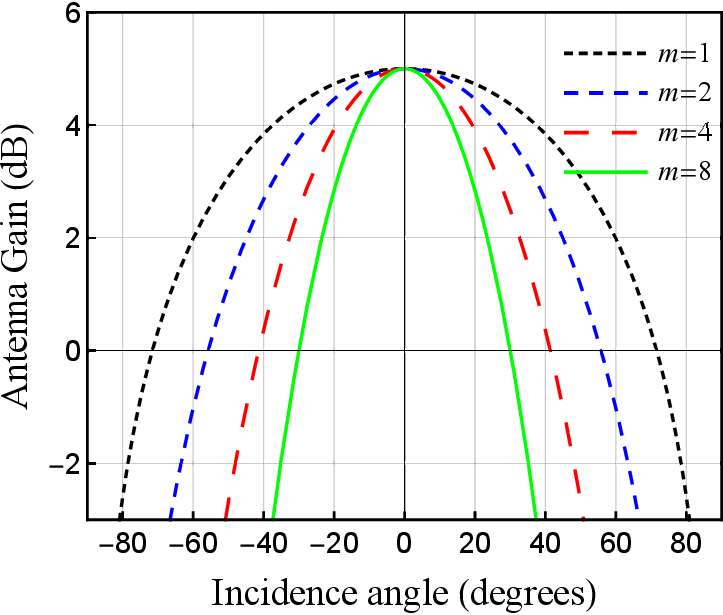}
\caption{Antenna gain vs. incidence angle for various values of $m$.}
\label{Figure3}
\end{figure}

Generally speaking, the gain of a directional antenna is a highly non-linear function of azimuth and elevation angles, adding considerable complexity to performance analysis. To simplify the analysis, the following gain is considered as described in \cite{J:Zhang}, \cite{J:Peng}.
\begin{equation}
G_{t}(\vartheta)=G_{0}\cos^{m}(\vartheta),
    \label{Gain}
\end{equation}
where $G_{0}$ represents the maximum antenna gain, $m$ is the directivity factor, which characterizes the shape of the beam, and $\vartheta\in[-\pi/2,\pi/2]$. Figure \ref{Figure3} illustrates the antenna gain (in dB) versus the incidence angle, $\vartheta$, for various values of the directivity factor, considering $G_{0}=5$dB.

The minimum SNR, $\gamma_{\text{min}}$, occurs at the boundary of the coverage area, where the path loss is maximized. Accurately determining this value requires accounting for the combined effects of the gains of the UAV and the Rx antenna. By incorporating these factors, a more complete characterization of the quality of the received signal can be obtained. This computation (in dB) can be expressed as follows
\begin{equation}
\gamma_{\text{min}}(H)=P_{t}+G_{t}(\theta)+G_{r}-PL_{\text{max}}(H)-P_{n}~~~(\text{in dBs}),
    \label{SNR}
\end{equation}
where $P_{t}$ is the transmit power of the UAV in dBm, $G_{t}(\theta), G_{r}$ are the antenna gains of the UAV and Rx, respectively, expressed in dB, $\theta$ is defined in \eqref{theta}, $PL_{\text{max}}$ is given by \eqref{PLmax}, and $P_{n}$ is the noise power in dBm. We further assume that the Rx is equipped with an omnidirectional antenna of unit gain, thereby minimizing its impact on the analysis. This assumption facilitates a more generalized performance evaluation without introducing bias into the results.

In this scenario, the optimal altitude, $H_{\text{OPT*}}$, is obtained by determining the root of the derivative of \eqref{SNR} with respect to $H$
\begin{equation}
\frac{\partial \gamma_{\text{min}}(H)}{\partial H}=0.
    \label{Der2}
\end{equation}
The resulting nonlinear equation, similar to \eqref{Eq1}, does not have a closed-form solution and must also be solved using numerical methods. 

\subsection{Optimal altitude vs. energy consumption}
The operational efficiency of UAVs is inherently constrained by their limited battery capacity, making energy consumption a pivotal factor in the optimization process. To systematically address this constraint, it is essential to analyze the key parameters influencing UAV power consumption in various flight phases.

In the following, the UAV is initially positioned at the intersection of the ellipse semi-axes. It ascends to the target altitude $H$ via vertical take-off and subsequently transitions to forward-level flight, maintaining a constant altitude while traveling toward its designated location. Upon reaching its destination, the UAV hovers to initiate the downlink data transmission to the Rx. During this hovering phase, a steady-state snapshot is examined assuming that both the UAV and the Rx remain stationary. The power consumption for hovering, forward flight, and vertical take-off is mathematically expressed as \cite{J:You}, \cite{J:Zeng}.
\begin{equation}
p_{\text{hov}}=\overbrace{\frac{\delta}{8}\varrho \varsigma A_{U} \rm{u}_{\text{tip}}^{3}}^{Z_{1}}+\overbrace{(1+\rm{k})\frac{\textit{N}_{\textit{U}}^{3/2}}{\sqrt{2\varrho \textit{A}_{\textit{U}} }}}^{Z_{2}},
    \label{phov}
\end{equation}
\begin{align}
p_{\text{for}}=&Z_{1}\left(1+\frac{3\rm{u}^2}{\rm{u}_{\text{tip}}^{2}} \right)+Z_{2}\left(\sqrt{1+\frac{\rm{u}^4}{4\rm{u}_{0}^4}}-\frac{\rm{u}^2}{2\rm{u}_{0}^2} \right)^{1/2} \nonumber \\ &+\frac{1}{2}\mathscr{D}\varrho\varsigma \textit{A}_{\textit{U}} \rm{u}^3,
    \label{pfor}
\end{align}

\begin{equation}
p_{\text{vto}}=Z_{1}+\frac{\textit{N}_{\textit{U}}\rm{u}_\text{to}}{2}+\frac{\textit{N}_{\textit{U}}}{2}\sqrt{\rm{u}_\text{to}^2+\frac{2\textit{N}_\textit{U}}{\varrho \textit{A}_\textit{U}}},
    \label{pvto}
\end{equation}
where $Z_{1}$ and $Z_{2}$ represent the power of the blade profile and the power induced during hovering, respectively. The parameters $\delta$, $\varrho$, and $\varsigma$ correspond to the drag coefficient of the profile, the density of air, and the solidity of the rotor. Furthermore, $\textit{A}_\textit{U}$, $\rm{u}_{\text{tip}}$, $\rm{k}$, and $\textit{N}_\textit{U}$ denote the area of the rotor disc, the tip speed of the rotor blade, the incremental correction factor for the induced power, and the weight of the UAV in Newtons, respectively. Furthermore, $\rm{u}$, $\rm{u}_{0}$, $\mathscr{D}$, and $\rm{u}_{\text{to}}$ represent the forward-level flight speed, the mean rotor-induced velocity in hover, the fuselage drag ratio, and the vertical takeoff speed, respectively.

Based on the above, the total energy consumption ($E_{C}$), measured in Joules, required to complete the data transmission from the downlink at the coverage boundary can be expressed as

\begin{align}
    E_{C}(H)=&\frac{p_{\text{vto}}H}{\rm{u}_{\text{to}}}+\frac{p_{\text{for}}\sqrt{H^2b^2+(b^2+H^2)(a^2-b^2)}}{\mathrm{u} b^2} \nonumber\\
    &+\left(p_{\text{hov}}+P_{t}\right)\frac{\mathcal{Q}}{\mathscr{W}\log_{2}(1+\gamma_{\text{min}})},
    \label{TEC}
\end{align}
where $\mathscr{W}$ represents the channel bandwidth and $\mathcal{Q}$ denotes the throughput requirement (in bits) that the UAV must deliver to Rx.

The optimal altitude, $H_{\text{OPT**}}$, corresponding to $E_{C}$, is determined by numerically solving the root of

\begin{equation} 
\frac{\partial E_{C}(H)}{\partial H} = 0. \label{Der3}
\end{equation}

This altitude signifies the point where energy consumption is minimized under the given conditions. The solution represents a critical trade-off between coverage and power consumption.

\section{Case Study}
To facilitate a clearer understanding of the procedures outlined above, we consider a typical quadrilateral $Q$ defined by the vertices $P_1 = (-200, -100)$, $P_2 = (-150, 300)$, $P_3 = (150, 350)$, and $P_4 = (200, 30)$ in the $\{x, y\}$ plane (with coordinates in meters). Its area can be found by applying the shoelace formula as $S=126,000$m$^2$ \cite{Weisstein}. Furthermore, specific values are selected for the parameter set $(\xi_{\text{LOS}}, \xi_{\text{NLOS}}, \eta, \kappa)$ based on different environments as follows: suburban $(0.1, 21, 4.88, 0.43)$, urban $(1, 20, 9.61, 0.16)$, dense urban $(1.6, 23, 12.08, 0.11)$, and high-rise urban $(2.3, 34, 27.23, 0.08)$ \cite{J:Al-Hourani}. In addition, the UAV is assumed to operate at a frequency of $f = 2 \, \text{GHz}$, and is characterized by the following set of parameters \cite{J:You}: $\delta=0.012$, $\varrho=1.225$kg/m$^{3}$, $\varsigma=0.05$, $\text{A}_{\text{U}}=0.503$m$^2$, $\text{u}_{\text{tip}}=120$m/sec, $\rm{k}=0.1$, $\text{N}_{\text{U}}=20$N, $\text{u}_{0}=4.03$m/sec, $\mathscr{D}=0.6$, $\text{u}=20$m/sec, $\text{u}_{\text{to}}=3$m/sec and $\mathscr{W}=1$MHz. Finally, the Tx and noise power are set to $20$dBm and $-120$dBm, respectively.

\subsection{Inscribed ellipse with largest area}
After estimating the parameters of the affine transformation $\mathbf{T_a}$ that maps $Q$ to $Q'$, we obtain 
\begin{equation}
\begin{array}{l}
x' = \frac{1}{{1,535}}\left(4x-0.5y+750 \right),\\ \\
y' = \frac{1}{{1,535}}\left(-1.3x-4y+140 \right),
\end{array}
\label{Aff2}
\end{equation}
The coordinates of the transformed vertex $P'_{3}$ are obtained by applying \eqref{Aff2} to the coordinates of the vertex $P_{3}$, resulting in $(s, t) = \left( \frac{235}{307}, \frac{269}{307} \right)$. The value of $q$ that maximizes the inscribed ellipse is determined from \eqref{qval} as $q = 0.527$. Subsequently, the $\mathcal{B}_i$ coefficients of the unique maximal-area ellipse are derived by applying the inverse affine transformation $\mathbf{T_{a}}^{-1}$ to \eqref{InscEl}, and are presented in Table I. The values of the major and minor semi-axes are determined from \eqref{alpha} and \eqref{beta}, respectively, as $(a, b) = (200.3, 155.2)$. The area is approximately $S=\pi ab=97,611.80$m$^2$ \cite[eq. (3.328a)]{B:Bronshtein}, resulting in a $77.47\%$ coverage.

\subsection{Circumscribed ellipse with smallest area}
The similarity transformation mapping $Q$ to $Q''$, can be written as
\begin{equation}
\begin{array}{l}
x'' = \frac{1}{1,625}\left(4x - 0.5y + 750\right),\\ \\
y'' = \frac{1}{1,625}\left(0.5x + 4y + 500\right).
\end{array}
\label{Iso2}
\end{equation}
Applying \eqref{Iso2} to the coordinates of the vertices $P_3$ and $P_4$, the transformed coordinates of the vertices $P''_3$ and $P''_4$ are obtained as $(s, t) = \left( \frac{47}{65}, \frac{79}{65} \right)$ and $(v, w) = \left( \frac{307}{325}, \frac{144}{325} \right)$, respectively. The value of $u$ that minimizes the circumscribed ellipse is determined from \eqref{uval} as $u = 1.610$. Subsequently, the $\mathcal{D}_i$ coefficients of the unique minimum-area ellipse are derived by applying the inverse similarity transformation $\mathbf{T}_s^{-1}$ to \eqref{CircEl} and are presented in Table I. The major and minor semi-axes are calculated from \eqref{alpha} and \eqref{beta}, as $(a, b) = (294.3, 223.5)$. The area is approximately $S = 206,536.80 \, \text{m}^2$, i.e., 39.99\% of the elliptic area falls outside $Q$.
\begin{table}[t]
\label{Table1}
\caption{Ellipse Coefficients.}
\centering
\renewcommand{\arraystretch}{1.2}
\begin{tabular}{l|c|c} 
  \hline \hline
  $i$& $\mathcal{B}_i$ & $\mathcal{D}_i$ \\ 
  \hline
  1 & $5.073\times 10^{-6}$ & $5.058\times 10^{-6}$ \\ 
  2 & $-2.291\times 10^{-6}$ & $-1.573\times 10^{-6}$ \\ 
  3 & $4.449\times 10^{-6}$ & $3.415\times 10^{-6}$ \\ 
  4 & $0.00033$ & $0.00027$ \\ 
  5 & $-0.00129$ & $-0.00075$ \\ 
  6 & $-0.04999$ & $-0.22675$ \\ 
    \hline \hline
\end{tabular}
\end{table}

\subsection{Optimal altitudes over all feasible inscribed and circumscribed ellipses}
\begin{figure}[h]
\centering
\subfigure[]{\includegraphics[width=0.45\columnwidth]{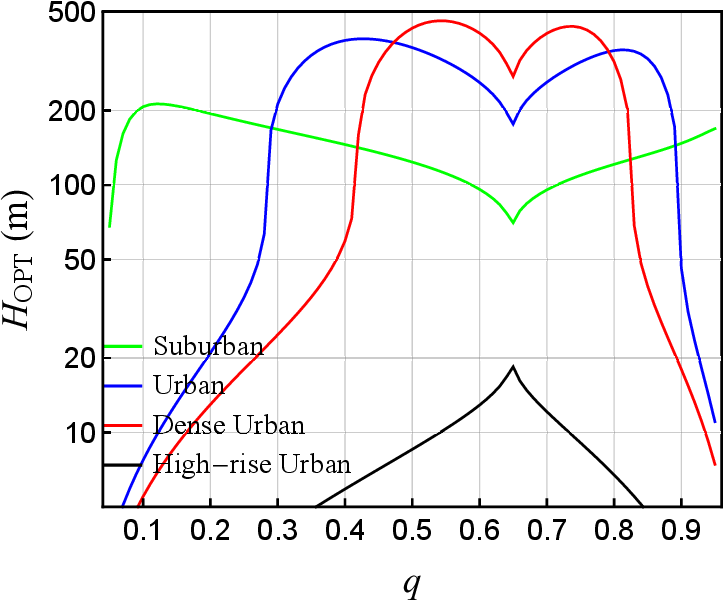}}
\subfigure[]{\includegraphics[width=0.45\columnwidth]{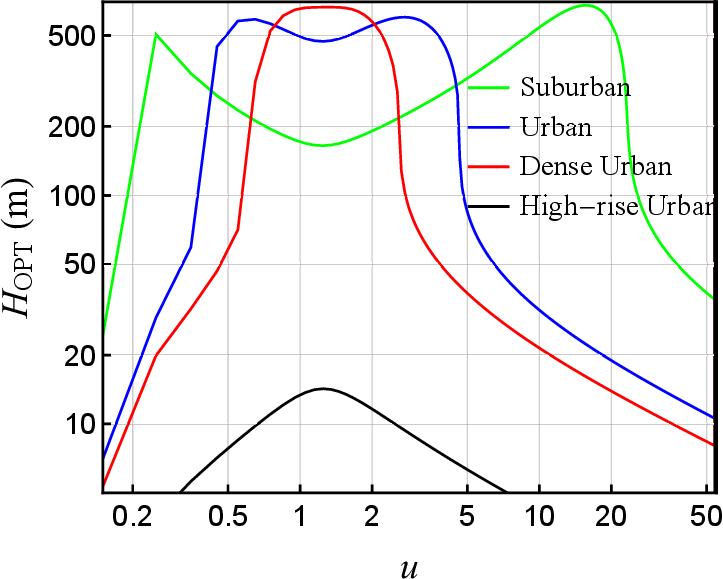}} 
\\
\subfigure[]{\includegraphics[width=0.45\columnwidth]{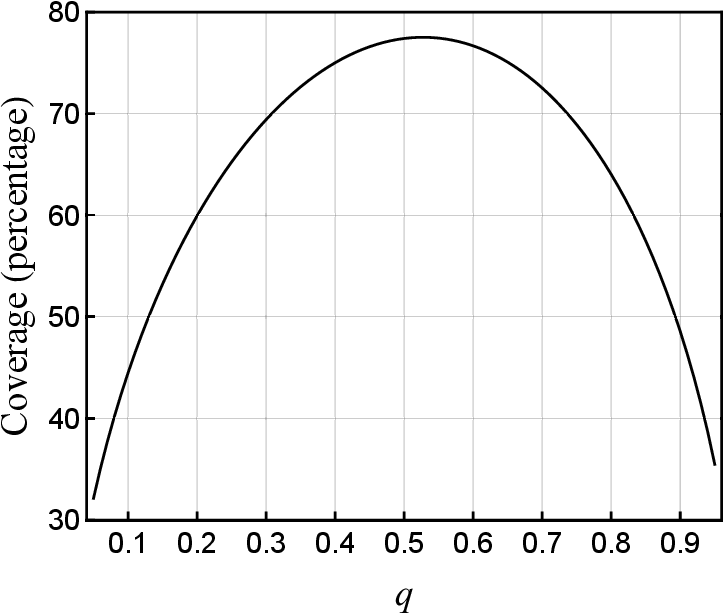}} 
\subfigure[]{\includegraphics[width=0.45\columnwidth]{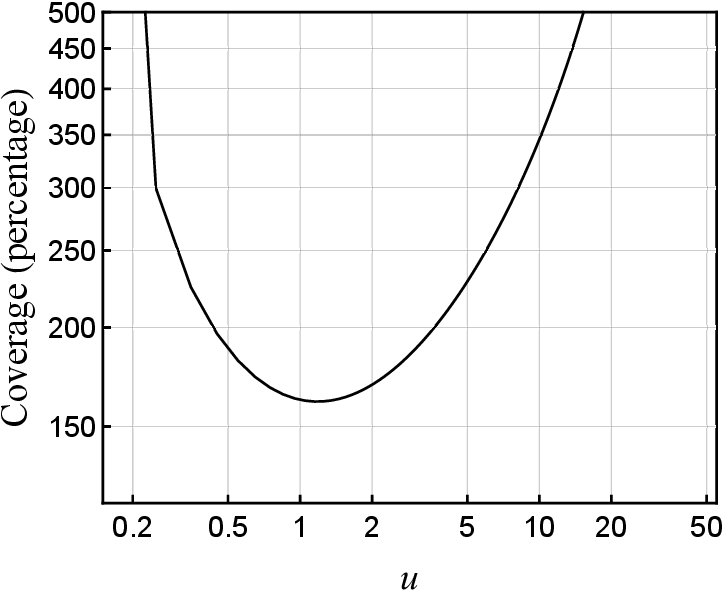}} 
\\
\subfigure[]{\includegraphics[width=0.45\columnwidth]{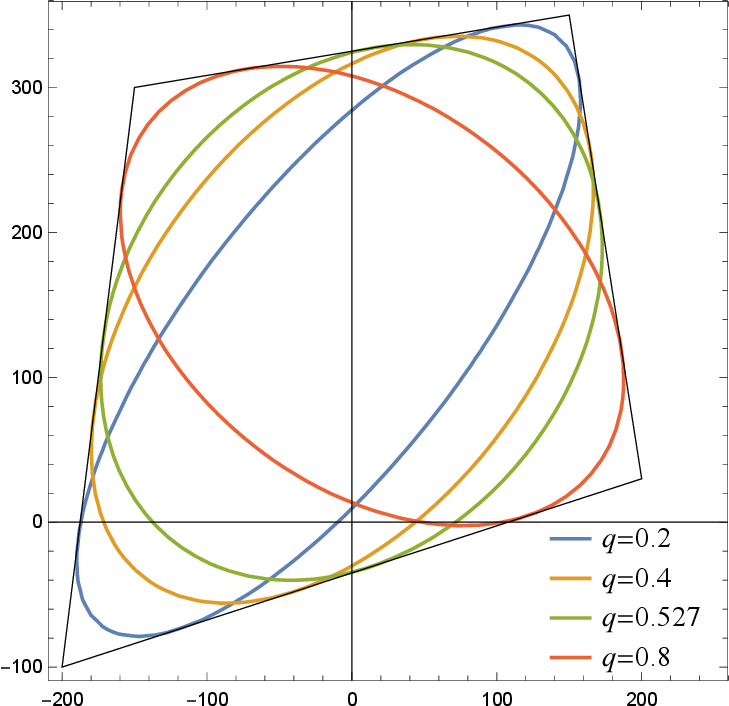}}
\subfigure[]{\includegraphics[width=0.45\columnwidth]{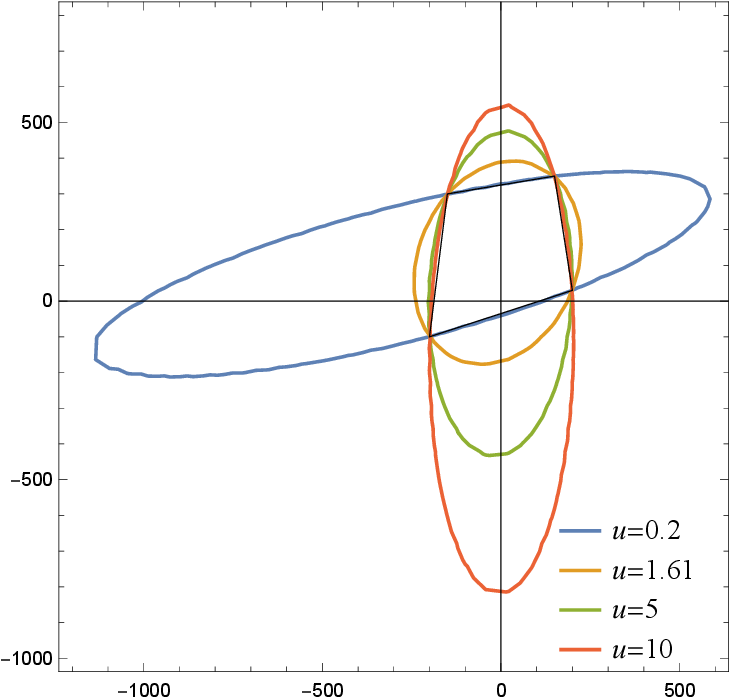}} 
\caption{Optimal altitude $H_{\mathrm{OPT}}$ as function of (a) $q$ for inscribed ellipses and (b) $u$ for circumscribed ellipses. Coverage area as function of (c) $q$ for inscribed ellipses and (d) $u$ for circumscribed ellipses. Representative ellipse configurations are shown in (e) for the inscribed case ($q = 0.2$–$0.8$) and in (f) for the circumscribed case ($u = 0.2$–$10$).}
\label{Figure4}
\end{figure}

The optimal altitudes with respect to the path loss are evaluated by determining the corresponding values for all feasible inscribed and circumscribed ellipses. Figure~\ref{Figure4} presents the optimal altitude $H_{\mathrm{OPT}}$ and the corresponding coverage for each propagation environment, evaluated over varying $q$ (inscribed ellipses, (a),(c),(e)) and $u$ (circumscribed ellipses, (b),(d),(f))\footnote{For the circumscribed ellipse, coverage is defined as the ratio of the elliptical area to the quadrilateral area.}. These results further validate the accuracy and robustness of the proposed methodology, as $H_{\mathrm{OPT}}$ is consistently obtained for all examined values of $q$ and $u$.

It is observed that, for inscribed ellipses, the feasible coverage area decreases significantly as $q$ deviates from its optimal value, whereas for circumscribed ellipses, the feasible coverage region increases considerably as $u$ departs from its corresponding optimum. Furthermore, the same methodology can be readily extended to scenarios involving SNR and energy consumption.

\subsection{Multiple quadrilaterals}
The proposed methodology is general and applicable to any convex quadrilateral, regardless of its shape, size, or orientation. To demonstrate its generality, results are provided for 1,000 quadrilaterals generated according to the following procedure. For clarity, only the inscribed ellipses are presented for the path loss case. A similar analysis, along with the corresponding results, can be readily extended to circumscribed ellipses, as well as to the cases of SNR and energy consumption.

\begin{figure}[h]
\centering

\subfigure[]{\includegraphics[width=0.49\columnwidth]{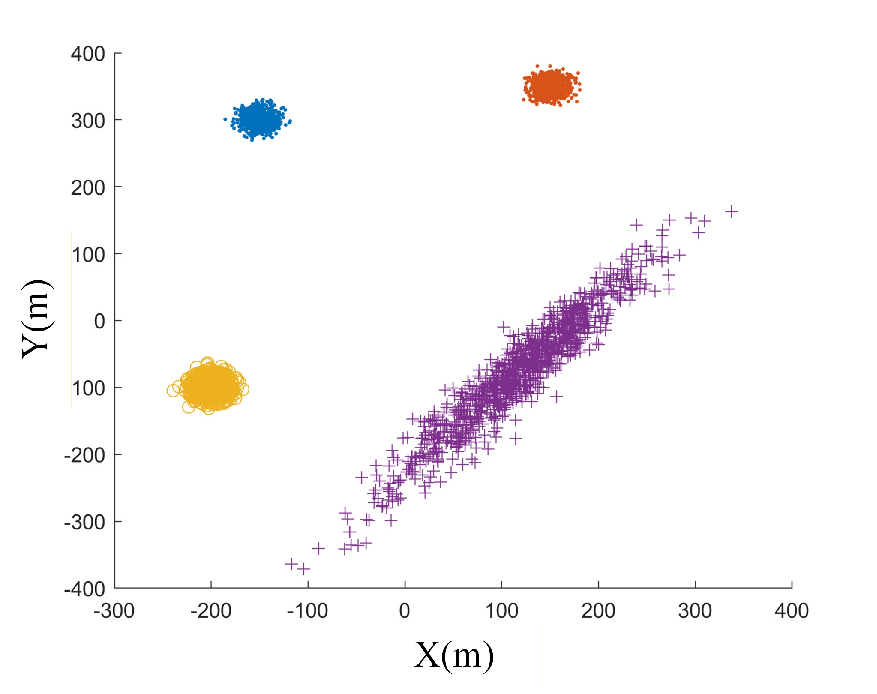}}
\subfigure[]{\includegraphics[width=0.49\columnwidth]{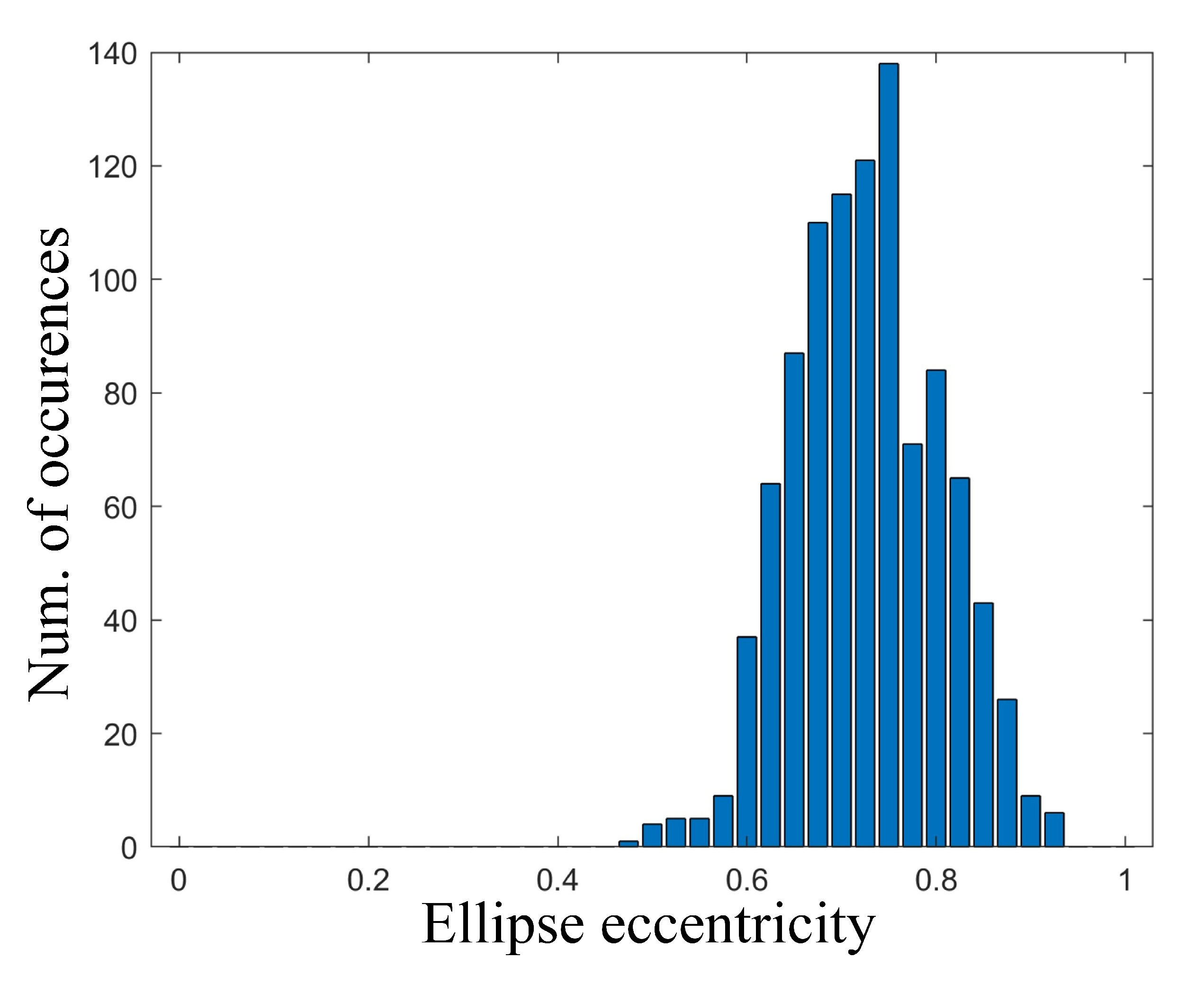}} \\
\subfigure[]{\includegraphics[width=0.49\columnwidth]{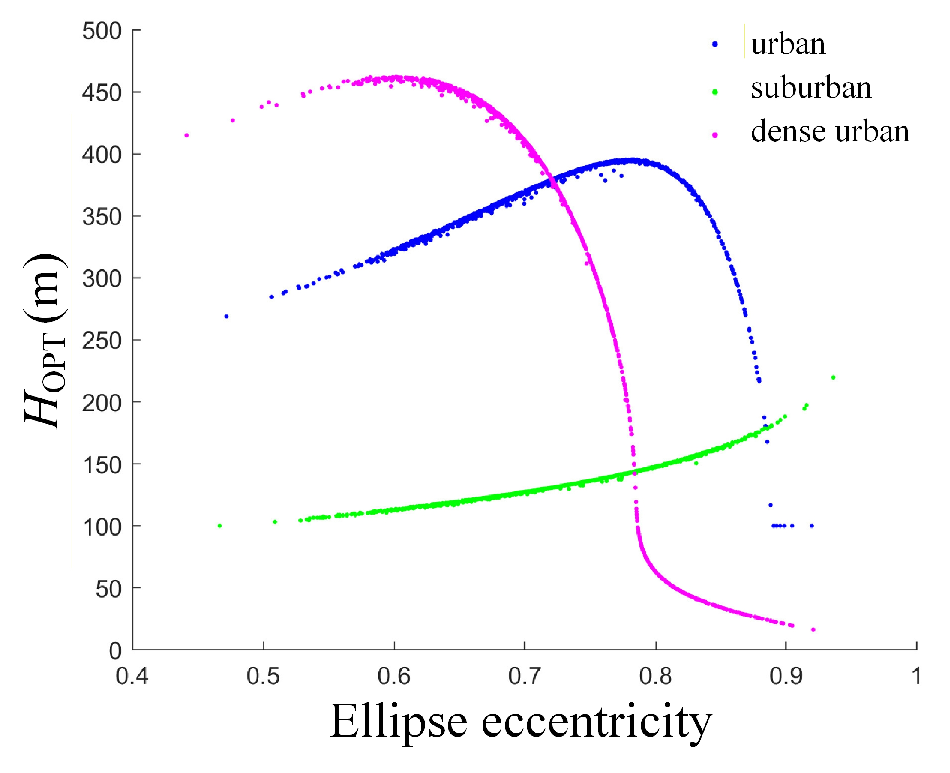}}
\subfigure[]{\includegraphics[width=0.49\columnwidth]{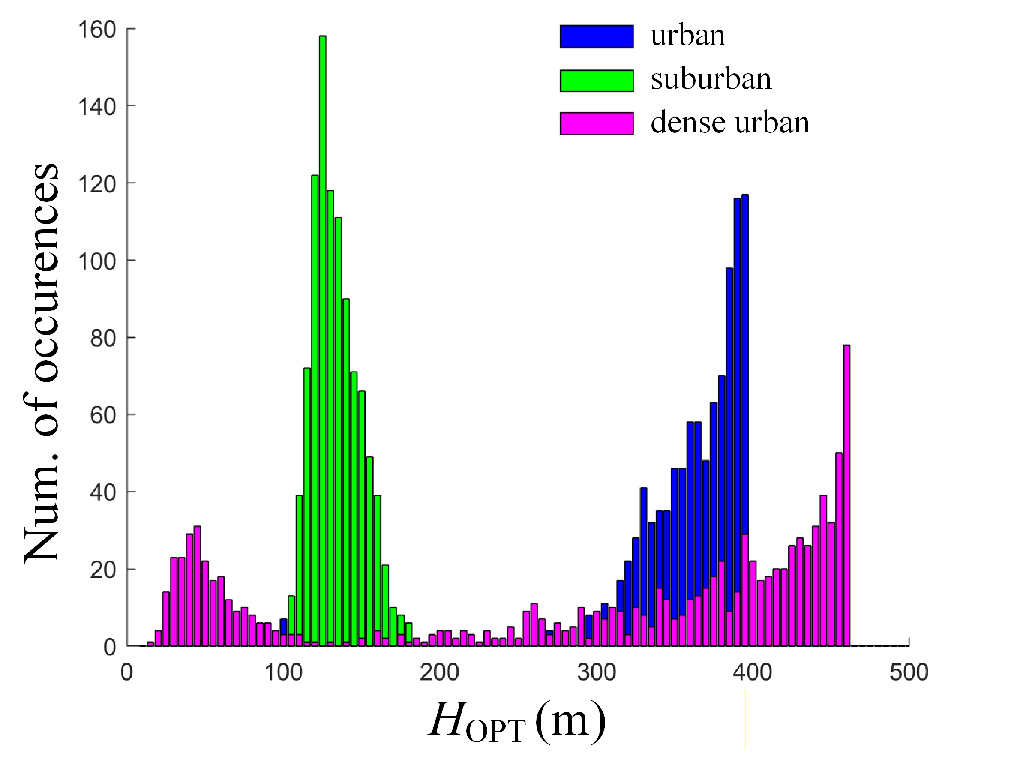}}
\caption{(a) Vertex scatter plot of 1,000 generated quadrilaterals, (b) histogram of ellipse eccentricities, (c) $H_{\mathrm{OPT}}$ vs. ellipse eccentricity, and (d) histogram of $H_{\mathrm{OPT}}$.}
\label{Figure5}
\end{figure}

Consider the original quadrilateral with vertices $P_1 -P_4$. Small random perturbations are applied to the first three vertices, $P_1, P_2, P_3$, by adding a random variable that follows a Gaussian distribution with zero mean and a standard deviation $\sigma=10$m to each coordinate. The position of the fourth vertex $P_4$ is then adjusted to ensure that the total area of the quadrilateral remains constant. 
Figure \ref{Figure5}(a) illustrates the resulting positions of vertices $P_1-P_4$, as obtained from the previously described procedure, for 1,000 generated quadrilaterals. Vertex $P_1$ corresponds to the yellow group, $P_2$ to the blue group, and $P_3$ to the red group, while $P_4$, calculated to preserve the area of the quadrilateral, corresponds to the purple group. The results indicate a wide diversity in shape and orientation while preserving convexity. Moreover, Fig. \ref{Figure5}(b) illustrates the histogram of ellipse eccentricities for the 1,000 generated quadrilaterals. The distribution indicates a wide range of ellipse shapes, predominantly elongated, with none approximating a circular disk (i.e., eccentricity values near 0).

Figure \ref{Figure5}(c) illustrates the obtained $H_{\mathrm{OPT}}$ as a function of ellipse eccentricity for the 1,000 quadrilaterals. As the propagation environment becomes more hostile, the value of $H_{\mathrm{OPT}}$ decreases for ellipses with higher eccentricities due to the increased distance of the farthest point and the increased maximum loss. Finally, histograms of $H_{\mathrm{OPT}}$ values, obtained using the proposed framework, are generated for the three propagation environments—suburban, urban, and dense urban—and presented in Fig. \ref{Figure5}(d).

\subsection{Application to a polygon partitioning}

In this subsection, we present typical examples that illustrate the application of the proposed method to the coverage of more general polygonal regions using elliptic footprints. We first consider the set of 1,000 randomly generated quadrilaterals introduced in the previous subsection and seek to cover each of them with four elliptical footprints. For this purpose, an interior point is selected within each quadrilateral and used to partition it into four smaller quadrilaterals. The proposed method is then applied independently to each sub-quadrilateral in order to determine the corresponding optimal inscribed elliptic footprint.

\begin{figure}[h]
\centering
\subfigure[]{\includegraphics[width=0.49\columnwidth]{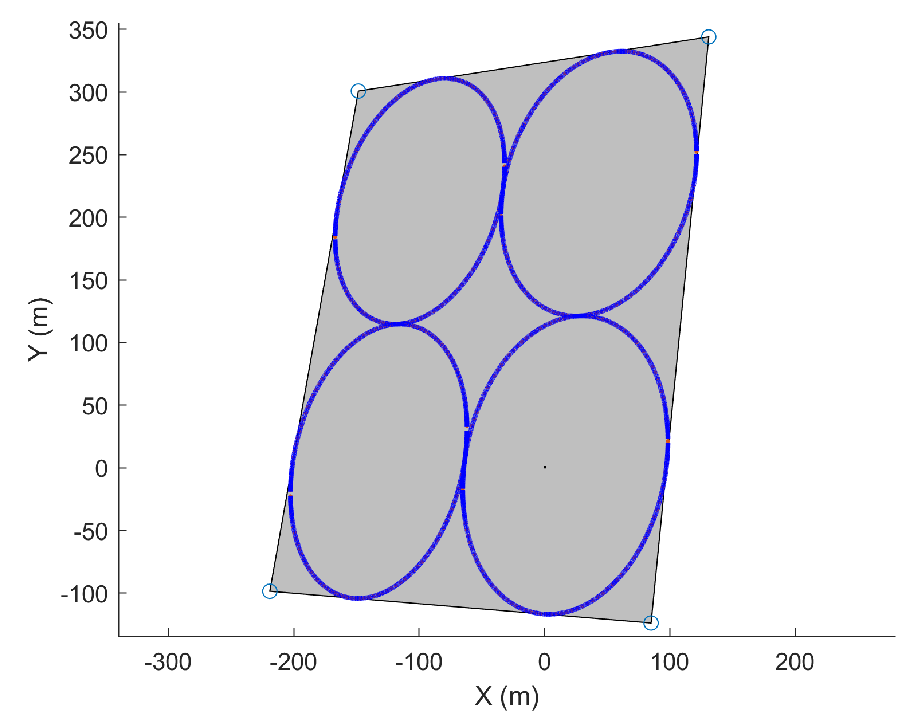}}
\subfigure[]{\includegraphics[width=0.49\columnwidth]{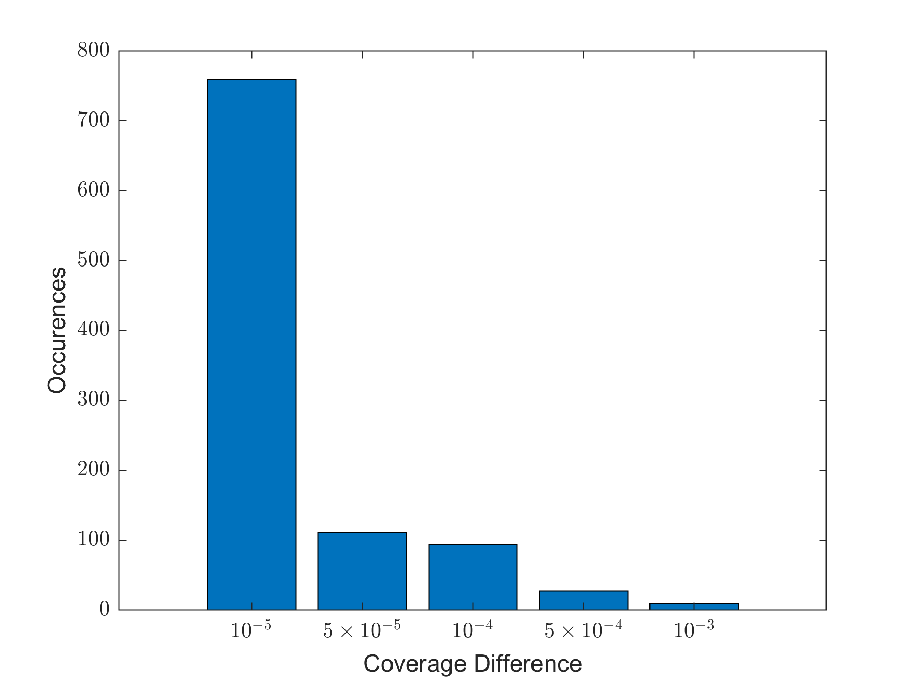}} \\
\caption{(a) Representative randomly generated quadrilateral from the 1,000 instances shown in Fig.~\ref{Figure5}, together with the four corresponding inscribed ellipses. (b) Histogram of the difference between the proposed inscribed coverage and the homography-based coverage method reported in~\cite{J:Vavoulas2}.}

\label{Figure6}
\end{figure}

To enable a direct comparison with the homography-based approach of \cite{J:Vavoulas2}, the partitioning point is initially determined using the homography-based method, and the proposed ellipse-construction method is subsequently applied to the resulting partition. Figure~\ref{Figure6}(a) illustrates the partitioning of a representative quadrilateral together with the corresponding inscribed ellipses. Figure~\ref{Figure6}(b) depicts the histogram of the coverage difference between the proposed method and the homography-based method, computed as the coverage achieved by the proposed inscribed ellipses minus that obtained by the homography-based approach.

As observed, the proposed method achieves higher coverage in all considered cases, although the improvement remains marginal. In most instances, the gain is on the order of $10^{-5}$, while the maximum observed improvement reaches $10^{-3}$. This behavior is expected since the partitioning is inherited from the homography-based method, whereas the proposed approach guarantees the maximum-area inscribed ellipse for each resulting quadrilateral.

In a second example, we consider the coverage of a regular hexagon using eight elliptical footprints. The hexagon is first decomposed into two identical quadrilaterals, and the homography-based method of \cite{J:Vavoulas2} is applied to each of them. This procedure achieves a total coverage of $0.7716$. Subsequently, the proposed inscribed-ellipse method is applied. In this case, each of the two quadrilaterals is further partitioned into four sub-quadrilaterals. The partitioning point is initialized near the intersection of the line segments connecting the midpoints of opposite edges. A constrained optimization procedure is then employed to determine the optimal partitioning point, using this intersection as the initial estimate. The algorithm converges after 10 iterations to a point very close to the initial estimate, yielding an improved coverage of $0.7741$.

\begin{figure}[h]
\centering
{\includegraphics[width=0.75\columnwidth]{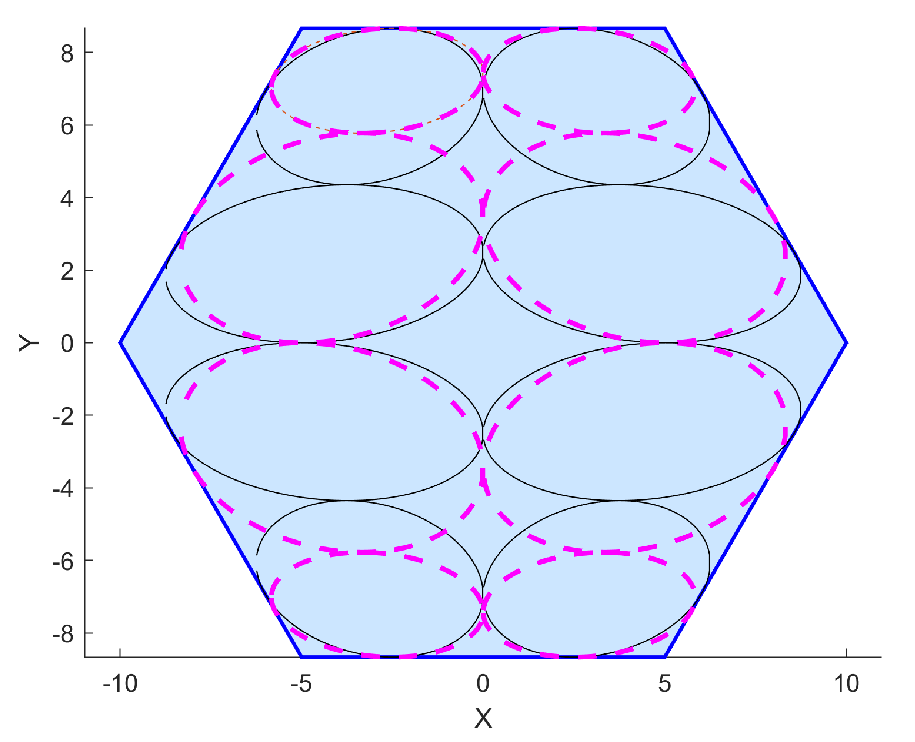}} 
\caption{Coverage of a regular hexagon using eight ellipses, comparing the homography-based approach (dashed contours) with the proposed inscribed-ellipse approach (solid contours).}

\label{Figure7}
\end{figure}

The use of the proposed optimal inscribed ellipse for quadrilateral regions as a building block for polygon coverage with elliptic footprints offers several advantages over purely heuristic optimization approaches. First, the coverage within each quadrilateral is optimal by construction. Second, the ellipse determination is deterministic and does not require computationally intensive iterative procedures. Third, geometric constraints, such as avoiding intersections between ellipses and preventing leakage outside the polygonal domain, are inherently satisfied. Finally, the resulting elliptic footprints have comparable areas, which is a desirable property in UAV coverage applications. Figure \ref{Figure7} demonstrates the regular hexagon and the elliptic coverage using homography (dashed ellipses) and the proposed inscribed ellipses (black continuous curves).

For comparison, the algorithm proposed in \cite{J:Kampas} addresses the same problem for regular polygons and reports a coverage of $0.82$. However, this higher coverage is obtained at the expense of marginal violations of the aforementioned geometric constraints. Moreover, the resulting ellipses exhibit highly non-uniform sizes, with one dominant ellipse and several substantially smaller ones, which may render the solution impractical for UAV coverage scenarios. In addition, the method is computationally intensive, with reported execution times of approximately $140\,\mathrm{s}$, thereby limiting its applicability in real-time settings. Finally, the method is designed specifically for regular polygons and cannot handle other realistic polygonal shapes.

It is also worth noting that, even when the optimization step is omitted and the midpoint-intersection point is used directly as the partitioning point, the coverage achieved by the proposed inscribed-ellipse construction is only marginally reduced, reaching approximately $0.76$.

\subsection{Discussion}

\begin{figure}[h]
\centering
\includegraphics[keepaspectratio,width=\columnwidth]{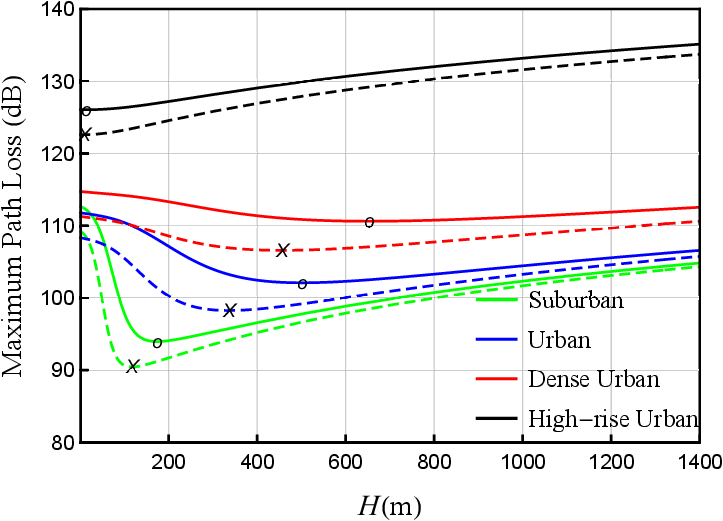}
\caption{Maximum path loss vs. altitude, with $H_{\text{OPT}}$ values highlighted for the inscribed (dashed lines) and the circumscribed (solid lines) ellipse.}
\label{Figure8}
\end{figure}

\begin{figure}[h]
\centering
\includegraphics[keepaspectratio,width=\columnwidth]{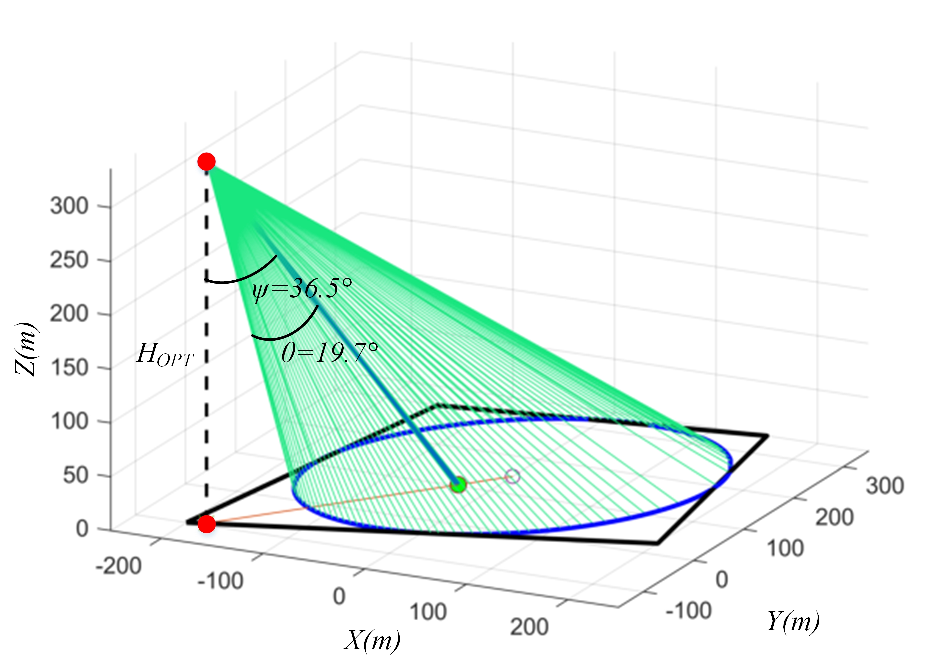}
\caption{Setup for covering the inscribed ellipse in an urban environment.}
\label{Figure9}
\end{figure}

Figure \ref{Figure8} illustrates the maximum path loss, $PL_{\text{max}}$, as a function of $H$ for various environments, considering the inscribed (dashed-line) and the circumscribed (solid-line) elliptical regions. In each scenario, the value of $H_{\text{OPT}}$ is indicated appropriately, and the required angles $\theta$ and $\psi$ are calculated using \eqref{psi} and \eqref{theta}, respectively. The corresponding values are summarized in Table \ref{Table2}. 
As an example, Fig. \ref{Figure9} visualizes the geometric set-up where the UAV covers the inscribed ellipse of an urban environment. As expected, the optimum altitude is higher in the circumscribed scenario because the UAV needs to cover a larger area.  The feasibility of the resulting scenarios can be easily assessed. For example, the impractical value of $\theta=85.5^0$ in a high-rise urban environment renders the desired area coverage with a single UAV unfeasible.

\begin{table}[h]
\centering
\caption{Optimal Altitude.}
\label{Table2}
\renewcommand{\arraystretch}{1.4}
\begin{tabular}{l|l|c|c|c}
\hline \hline
 & Environment & $H_{\text{OPT}}$ (m) & $\theta$ ($^\circ$) & $\psi$ ($^\circ$) \\ 
\hline
\multirow{4}{*}{\rotatebox{90}{Inscribed}} & Suburban & 116.9 & 45.8 & 26.1 \\ 
 & Urban & 335.8 & 19.7 & 36.5 \\ 
 & Dense Urban & 456.0 & 14.8 & 37.7 \\ 
 & High-rise Urban & 9.5 & 85.5 & 2.8 \\ 
\hline
\multirow{4}{*}{\rotatebox{90}{Circumscribed}} & Suburban & 173.7 & 44.3 & 27.3 \\
 & Urban & 501.3 & 18.7 & 38.0 \\ 
 & Dense Urban & 653.3 & 14.6 & 39.0 \\ 
 & High-rise Urban & 13.3 & 85.5 & 2.9 \\ 
\hline \hline
\end{tabular}
\end{table}

\begin{figure}[h]
\centering
\includegraphics[keepaspectratio,width=\columnwidth]{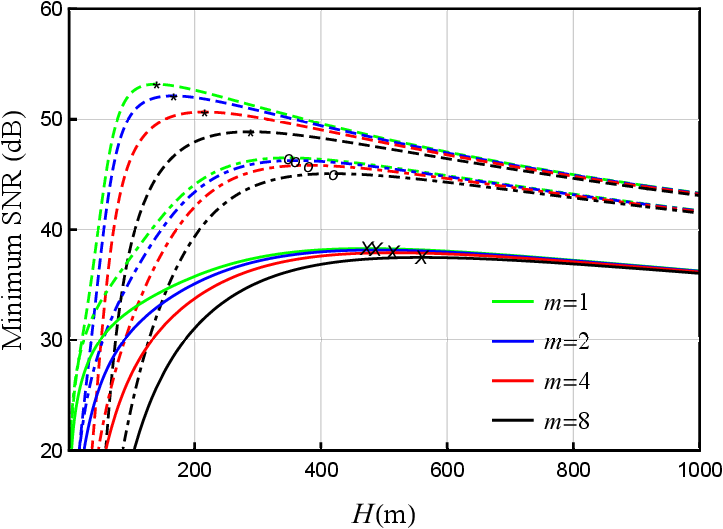}
\caption{Minimum SNR as a function of altitude, with $H_{\text{OPT}}$ values indicated for the inscribed ellipse in suburban (dashed lines), urban (dot-dashed lines), and dense urban (solid lines) environments.}
\label{Figure10}
\end{figure}

\begin{figure}[h]
\centering
\includegraphics[keepaspectratio,width=\columnwidth]{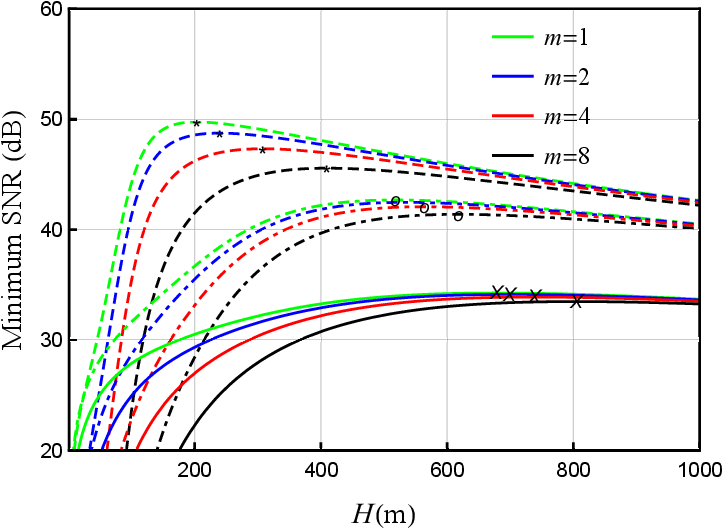}
\caption{Minimum SNR as a function of altitude, with $H_{\text{OPT*}}$ values indicated for the circumscribed ellipse in suburban (dashed lines), urban (dot-dashed lines), and dense urban (solid lines) environments.}
\label{Figure11}
\end{figure}

Figures \ref{Figure10} and \ref{Figure11} demonstrate the minimum SNR, $\gamma_{\text{min}}$, as a function of $H$ for various values of the directivity factor $m$, considering the inscribed and circumscribed elliptical regions, respectively. The results indicate that as $m$ increases, leading to a more directional antenna, the optimal altitude, $H_{\text{OPT*}}$, increases while the corresponding minimum SNR decreases. This is attributed to the narrower beamwidth, which reduces received power at the coverage boundary despite improving gain along the main lobe. Furthermore, in dense urban settings, the steep decrease in SNR at lower altitudes highlights the impact of severe NLoS conditions, which requires a higher positioning of UAVs to mitigate excessive path loss.

\begin{figure}[h]
\centering
\includegraphics[keepaspectratio,width=\columnwidth]{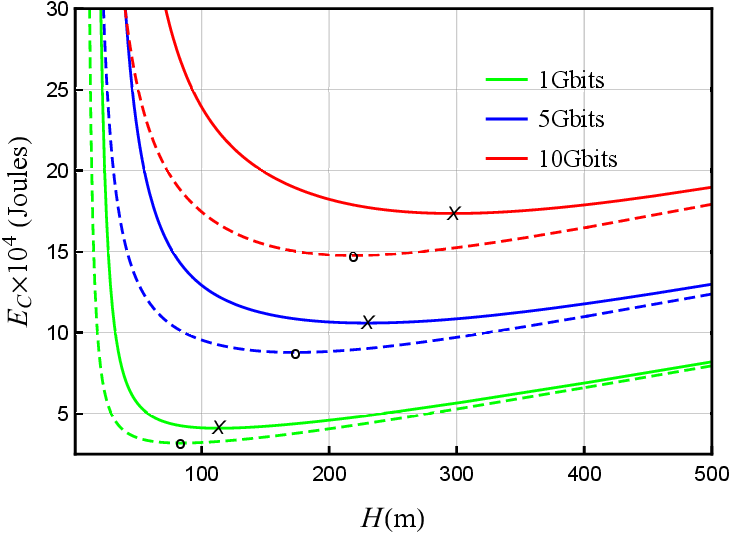}
\caption{Total energy consumption vs. altitude, with $H_{\text{OPT**}}$ values highlighted for the inscribed (dashed lines) and circumscribed (solid lines) ellipses.}
\label{Figure12}
\end{figure}

Finally, Fig. \ref{Figure12} illustrates the energy consumption of the UAV, $E_C$, as a function of $H$, for both the inscribed and circumscribed ellipses, considering different throughput requirements. The results demonstrate that energy consumption initially decreases with increasing altitude due to reduced aerodynamic drag in forward flight but shifts beyond a certain threshold as path loss increases transmission power demands. The optimal altitude, $H_{\text{OPT**}}$, is lower for higher throughput requirements, since maintaining a strong SNR at lower altitudes reduces the transmission duration and total energy expenditure. The comparison between the inscribed and circumscribed cases further highlights the trade-off between full coverage and energy efficiency, with the circumscribed scenario requiring higher energy consumption due to the UAV's need to operate at a higher altitude.

Although the nonlinear equations arising from the optimization of UAV altitude with respect to path loss, SNR, and energy consumption do not admit closed-form analytical solutions, their numerical evaluation provides meaningful insight into system performance. The results presented in Figures \ref{Figure8}, \ref{Figure10}, \ref{Figure11}, and \ref{Figure12} illustrate the inherent trade-offs among these key metrics, demonstrating how changes in UAV altitude can lead to varying coverage quality, energy efficiency, and communication reliability. This approach allows for the identification of optimal operating points under practical constraints, even in geometrically complex coverage regions where analytical solutions are not feasible. 

In general, the results emphasize the importance of selecting an appropriate altitude based on environmental conditions, antenna directivity, and energy constraints. The findings provide valuable information for the design of UAV-assisted communication networks that ensure optimal coverage while minimizing energy consumption.

\section{Conclusions and Further Research}
This paper investigated the optimal deployment of a single UAV to provide efficient coverage over an arbitrary convex quadrilateral region using an elliptical footprint. Two coverage scenarios were considered: (i) the largest inscribed ellipse, which ensures coverage within the quadrilateral while excluding its perimeter, and (ii) the smallest circumscribed ellipse, which guarantees full region coverage. A comprehensive optimization framework was developed to determine the optimal UAV altitude, incorporating key performance metrics such as path loss, SNR, and energy consumption.

The study revealed that the optimal altitude of UAVs is highly dependent on environmental conditions, antenna directivity, and communication constraints. It was found that increasing antenna directivity leads to a higher optimal altitude while simultaneously reducing the minimum required SNR. Furthermore, the analysis of UAV energy consumption underscored the trade-off between maximizing coverage and minimizing power usage, highlighting the critical need to balance flight dynamics with communication performance.  In particular, the analysis over all feasible ellipse configurations and the large-scale evaluation over randomly generated quadrilaterals confirm the robustness and generality of the proposed framework. These findings offer valuable insights for the design of energy-efficient UAV-assisted communication networks, especially in urban and high-rise environments where NLoS conditions significantly affect performance.

To account for potential model mismatch and enhance practical relevance, future work will incorporate more realistic channel models, including stochastic geometry-based formulations and environment-aware LoS probability models with spatial uncertainty. In addition, sensitivity analysis with respect to key propagation parameters will be conducted to quantify the robustness of the optimal altitude under varying environmental conditions. This will provide further insight into the reliability of the proposed framework for real-world deployment scenarios.

In our recent work \cite{J:Vavoulas2}, seamless multi-UAV coverage is achieved by coordinating elliptical footprints through deterministic packing and homography transformations, ensuring tangency and gap-free deployment. As future work, coordination strategies based on region partitioning can be further investigated to enable efficient and scalable multi-UAV coverage. The integration of UAV-mounted movable antenna (MA) technologies into the proposed framework, enabling joint optimization of UAV altitude, antenna configuration, and coverage footprint, is also a promising direction. In addition, adaptive antenna strategies and their impact on user performance and system-level trade-offs can be studied. Finally, issues such as inter-UAV interference, coordination overhead, and error propagation across subregions will be systematically analyzed. This extension will provide a more comprehensive validation of the scalability and practical applicability of the proposed framework.

\appendices
\renewcommand{\thesection}{A}
\section{Proof of unimodality of $PL_{\rm max}(H)$}
\setcounter{equation}{0} \renewcommand\theequation{A.\arabic{equation}}
\label{Unimodality}
The path loss function, $PL_{\rm max}(H)$, as a function of the UAV altitude $H$, is expressed as
\begin{align}
PL_{\rm max}(H) &= \underbrace{\frac{\xi_{\rm LoS} - \xi_{\rm NLoS}}{1 + \eta \exp \big(-\kappa (\arctan(\phi_1(H)) - \eta)\big)}}_{f_1(H)} \\&+ \underbrace{10 \log_{10} \Big(H^2 + \phi_2(H)^2\Big)}_{f_2(H)} \nonumber \\&+ \text{Constants} \nonumber ,
\end{align}
where
\begin{align}
    \phi_1(H) = \frac{H b}{a b + \sqrt{(b^2 + H^2)(a^2 - b^2)}}, \\
\phi_2(H) = \frac{a b + \sqrt{(b^2 + H^2)(a^2 - b^2)}}{b}.
\label{phi12}
\end{align}
Keeping in mind that $C_{1}=\xi_{\rm LoS}-\xi_{\rm NLoS}<0$, we define $\mathbf{d}=\sqrt{a^2-b^2}$, $\mathbf{r}(H)=\sqrt{H^2+b^2}$, $\mathbf{t}(H)=\arctan\big(\phi_1(H)\big)$, and $\mathbf{q}(H)=\eta \exp\big(-\kappa(\mathbf{t}(H)-\eta)\big)>0$. Then the auxiliary functions in \eqref{phi12} can be rewritten as
\begin{equation}
\phi_2(H)=a+\frac{\mathbf{d}}{b}\mathbf{r}(H),
\qquad
\phi_1(H)=\frac{H}{\phi_2(H)}.
\end{equation}
Furthermore, it can be readily shown through straightforward algebraic manipulations that:
\begin{equation}
H^2+\phi_2(H)^2
=
\left(\frac{a}{b}\mathbf{r}(H)+\mathbf{d}\right)^2.
\label{eq:Hphi2_identity}
\end{equation}
Hence,
\begin{equation}
f_2(H)=20\log_{10}\left(\frac{a}{b}\mathbf{r}(H)+\mathbf{d}\right),
\end{equation}
and therefore
\begin{equation}
f_2'(H)
=
\frac{20a}{\ln (10)}\,
\frac{H}{\mathbf{r}(H)\big(a\mathbf{r}(H)+b\mathbf{d}\big)}.
\label{eq:f2prime}
\end{equation}

Using \(\phi_1(H)=H/\phi_2(H)\), one obtains after simplification
\begin{equation}
\mathbf{t}'(H)
=
\frac{b^2}{\mathbf{r}(H)\big(a\mathbf{r}(H)+b\mathbf{d}\big)}.
\label{eq:tprime}
\end{equation}
Then
\begin{equation}
f_1(H)=\frac{C_{1}}{1+\mathbf{q}(H)},
\end{equation}
and differentiation yields
\begin{equation}
f_1'(H)
=
C_{1}\,\kappa\,
\frac{\mathbf{q}(H)}{(1+\mathbf{q}(H))^2}\,\mathbf{t}'(H).
\label{eq:f1prime_general}
\end{equation}
Substituting \eqref{eq:tprime} into \eqref{eq:f1prime_general} gives
\begin{equation}
f_1'(H)
=
C_{1}\,\kappa\,
\frac{\mathbf{q}(H)}{(1+\mathbf{q}(H))^2}\,
\frac{b^2}{\mathbf{r}(H)\big(a\mathbf{r}(H)+b\mathbf{d}\big)}.
\label{eq:f1prime}
\end{equation}

Combining \eqref{eq:f2prime} and \eqref{eq:f1prime}, the derivative of \(PL_{\max}(H)\) can be written as
\begin{equation}
PL'_{\max}(H)
=
\frac{
\frac{20a}{\ln(10)}\,H
+
C_{1}\,\kappa\,b^2\,
\frac{\mathbf{q}(H)}{(1+\mathbf{q}(H))^2}
}{\mathbf{r}(H)\big(a\mathbf{r}(H)+b\mathbf{d}\big)}
.
\label{eq:PLprime_factorized}
\end{equation}
Since the denominator in \eqref{eq:PLprime_factorized} is strictly positive for all $H\ge 0$, the stationary points of $PL_{\max}(H)$ are the zeros of
\begin{equation}
\mathcal{N}(H)=
\frac{20a}{\ln(10)}\,H
+
C_{1}\,\kappa\,b^2\,
\frac{\mathbf{q}(H)}{(1+\mathbf{q}(H))^2}.
\label{eq:NH_definition}
\end{equation}

Therefore, a sufficient condition for uniqueness of the stationary point is that $\mathcal{N}(H)$ be strictly increasing on $H\ge 0$. Differentiating \eqref{eq:NH_definition} gives
\begin{equation}
\mathcal{N}'(H)=
\frac{20a}{\ln(10)}
+
C_{1}\,\kappa\,b^2\,
\frac{\mathrm{d}}{\mathrm{d}H}
\left(
\frac{\mathbf{q}(H)}{(1+\mathbf{q}(H))^2}
\right).
\end{equation}
Using
\begin{equation}
\mathbf{q}'(H)=-\kappa \mathbf{q}(H)\,\mathbf{t}'(H)
\end{equation}
and
\begin{equation}
\frac{\mathrm{d}}{\mathrm{d}\mathbf{q}}\left(\frac{\mathbf{q}}{(1+\mathbf{q})^2}\right)
=
\frac{1-\mathbf{q}}{(1+\mathbf{q})^3},
\end{equation}
we obtain
\begin{equation}
\mathcal{N}'(H)=
\frac{20a}{\ln(10)}
-
|C_{1}|\,\kappa^2 b^2\,
\mathbf{t}'(H)\,
\frac{\mathbf{q}(H)\,|\mathbf{q}(H)-1|}{(1+\mathbf{q}(H))^3}.
\label{eq:Nprime_bound_start}
\end{equation}
Now observe that for $\mathbf{q}>0$,
\begin{equation}
\frac{\mathbf{q}|\mathbf{q}-1|}{(1+\mathbf{q})^3}\le \frac{\sqrt{3}}{18},
\label{eq:q_bound}
\end{equation}
and from \eqref{eq:tprime},
\begin{equation}
\mathbf{t}'(H)=\frac{b^2}{\mathbf{r}(H)\big(a\mathbf{r}(H)+b\mathbf{d}\big)}
\le
\frac{1}{a+\mathbf{d}},
\label{eq:tprime_bound}
\end{equation}
because $\mathbf{r}(H)\ge b$ for all $H\ge 0$. Combining \eqref{eq:Nprime_bound_start}--\eqref{eq:tprime_bound} yields
\begin{equation}
\mathcal{N}'(H)
\ge
\frac{20a}{\ln(10)}
-
|C_{1}|\,\kappa^2 b^2\,
\frac{\sqrt{3}}{18(a+\mathbf{d})}.
\end{equation}
Hence, a sufficient condition for $\mathcal{N}'(H)>0$ for all $H\ge 0$ is
\begin{equation}
\frac{20a}{\ln(10)}
>
|C_{1}|\,\kappa^2 b^2\,
\frac{\sqrt{3}}{18\left(a+\sqrt{a^2-b^2}\right)}.
\label{eq:unimodality_condition}
\end{equation}

Under \eqref{eq:unimodality_condition}, the function $\mathcal{N}(H)$ is strictly increasing, and thus
$PL'_{\max}(H)$ can vanish at most once. Since $PL'_{\max}(0)<0$ and $PL'_{\max}(H)>0$ for sufficiently large $H$, it follows that $PL_{\max}(H)$ has exactly one stationary point, which is its unique global minimum. Therefore, $PL_{\max}(H)$ is strictly unimodal on $H\ge 0$. It is straightforward to verify that the sufficient condition \eqref{eq:unimodality_condition} holds for all sets of parameters $\{a,b,\xi_{\rm LoS},\xi_{\rm NLoS},\kappa\}$. In particular, for the inscribed ellipse with $\{a,b\}=\{200.3,155.2\}$, the left-hand side evaluates to $1739.67$, while the right-hand side takes the values $\{38.26,4.82,2.56,2.01\}$ for suburban, urban, dense-urban, and high-rise urban scenarios, respectively, remaining significantly smaller in all cases. Similarly, for the circumscribed ellipse with $\{a,b\}=\{294.3,223.5\}$, the left-hand side evaluates to $2556.26$, whereas the right-hand side takes the values $\{38.24,4.81,2.56,2.01\}$, thus validating \eqref{eq:unimodality_condition}. Hence, $PL_{\max}(H)$ is strictly unimodal.

\bibliographystyle{IEEEtran}
\bibliography{IEEEabrv,References}

\balance
\end{document}